\documentclass[preprint,review,12pt]{elsarticle}

\usepackage{graphicx}
\usepackage{tikz}
\usetikzlibrary{decorations.pathreplacing}
\usetikzlibrary{decorations.pathmorphing,patterns}
\usepackage{amssymb}
\usepackage{amsmath}
\usepackage{stmaryrd}

\usepackage{lineno}
\usepackage{pdflscape}
\usepackage{subcaption}
\usepackage{multirow}

\newcommand*\diff{\mathop{}\!\mathrm{d}}

\journal{Journal of Computational Physics}

\begin{document}

\begin{frontmatter}



\title{Immersed Boundary Simulations of Flows Driven by Moving Thin Membranes}
\author{Marin Lauber\corref{cor1}\fnref{label1}}
\cortext[cor1]{Corresponding author}
\ead{M.Lauber@soton.ac.uk}
\author{Gabriel D. Weymouth\fnref{label1,label2}}
\address[label1]{Faculty of Engineering and Physical Sciences, University of Southampton, UK}
\address[label2]{Data-Centric Engineering Programme, Alan Turing Institute, UK}
\author{Georges Limbert\fnref{label1,label3}}
\address[label3]{Faculty of Health Sciences, University of Cape Town, South Africa.}


\begin{abstract}
Immersed boundary methods are extensively used for simulations of dynamic solid objects interacting with fluids due to their computational efficiency and modelling flexibility compared to body-fitted grid methods. However, thin geometries, such as shells and membranes, cause a violation of the boundary conditions across the surface for many immersed boundary projection algorithms. \textcolor{black}{Using a one-dimensional analytical derivation and multi-dimensional numerical simulations, this manuscript shows that adjustment of the Poisson matrix itself is require to avoid large velocity, pressure, and force prediction errors when the pressure jump across the interface is substantial and that these errors increase with Reynolds number. A new minimal thickness modification is developed for the Boundary Data Immersion Method (BDIM-$\sigma$), which avoids these issues while still enabling the use of efficient projection algorithms for high-speed immersed surface simulations.}
\end{abstract}


\end{frontmatter}


\section{Introduction}
\label{S:1}

Immersed boundary methods are a commonly used approach in computational fluid dynamics to simulate flow in complex domains or with moving or deforming boundaries~\cite{Kim2019,Griffith2020}. By solving the governing equations on a static Cartesian grid, mesh updates and/or re-meshing are completely avoided, rendering those methods highly efficient, at least from a computational point of view. However, imposing the correct boundary condition is less trivial than in body-fitted methods. Numerous methods have been developed to impose those boundary conditions on the body (see~\citep{Mittal2005} for a thorough review), and in the following, we will use the term \emph{immersed boundary method} to refer to these methods in general. The original Immersed Boundary method (IB) \cite{Peskin72} uses a regularized Dirac delta function forcing to spread the reaction force of the body (obtained from a constitutive law) from the Lagrangian points onto the fluid. While this approach has been successful at simulating fluid-structure interaction cases consisting of a flexible body dominated by the flow advection~\cite{Zhu2002}, the method can lead to poor mass conservation near the body \cite{Peskin1993}, resulting in a fluid leak across the interface \cite{griffith2012volume}. This leakage was identified as being the result of a divergent interpolated velocity field that drives the Lagrangian points. Ways of improving this volume conservation by constructing divergence-free interpolated velocity field have been derived, see~\citet{Bao2017}. 

The immersed boundary can also be considered in a purely discretized setting, where it is explicitly defined to enforce the appropriate boundary condition onto the body, as in the Direct-forcing method \cite{mohd1997combined,Fadlun2000}. The schemes that are used to reconstruct the velocity field in the vicinity of the immersed body are key to the accuracy of this method. For example, \citet{Fadlun2000} showed that volume-averaged forcing only results in $1^{st}$-order convergence of the velocity field, while linear interpolation improves the results with $2^{nd}$-order convergence. In addition to the velocity field adjustment, \textcolor{black}{Balaras \cite{Balaras2004} shows that Direct-forcing interpolation on a stationary body \textit{implicitly} imposes a consistent homogeneous Neumann boundary condition onto the pressure-correction.} However, immersed boundary methods are typically applied to moving geometries and unsteady simulations where the equations are solved using a fractional step algorithm~\cite{Chorin1967}, and in this case, the projection step introduces a slip error onto the velocity field \cite{Fadlun2000, Taira2007_1,GUY20102479,Kempe2012}\textcolor{black}{. In the case of thin shells, this error is significant  \cite{Gsell2021} and methods have been derived to reduce it, however, they are not able to completely remove it.}

Another class of immersed boundary methods adjust the pressure equation in addition to the velocity field during the fractional step update. The Boundary Data Immersion Method (BDIM) \cite{Weymouth2011,Maertens2015} explicitly adjusts the Poisson matrix such that the Neumann boundary condition on the pressure is respected on the immersed body. The resulting Poisson system has spatially varying coefficients, but it is otherwise unchanged. In these cases, standard linear algebra methods can be used to determine the pressure efficiently. Alternatively, the influence of the immersed boundary on the pressure field can be treated as an additional Lagrange multiplier as in the Immersed Boundary Projection Method (IBPM)\cite{Taira2007_1,Colonius2008}. Including this additional constraint into the Poisson equation results in an augmented system whose solution ensures imposition of the correct boundary condition on the velocity and pressure fields at the cost of solving a non-standard coupled Poisson problem.

In this work, we will show that when the immersed body is thin and dynamic, as is the case in many fluid-structure interaction problems (parachutes, insects wing, sail, heart valves, etc.), the lack of explicit boundary conditions on the pressure results in violation of both the pressure and velocity boundary conditions and therefore produces an erroneous solution. In cases where the pressure jump across the thin body is the dominant contribution to the overall force experienced by the body, the error is significant, and leads to large force prediction errors. Section~\ref{S:1D} will use a simple one-dimensional (1D) example to illustrate that correcting this error requires explicit modification of the Poisson projection step. \textcolor{black}{This careful analysis allows us to specify a set of mandatory numerical conditions for any Immersed Boundary method to successfully simulate problems involving thin dynamic surfaces}. Section~\ref{bdimsigma} will generalize this finding to 2D and 3D simulations of the Navier-Stokes equations, showing that the error increases with Reynolds number. In Section~\ref{sec:bdim-sigma} we develop a modified approach termed BDIM-$\sigma$ that allows for an efficient and accurate treatment of thin dynamic surfaces. Sections \ref{S:Convergence}, \ref{S:Disk} and \ref{S:Wing} verify and validate the new approach and compare against existing Immersed Boundary methods for 2D and 3D flows generated by thin membranes including an oscillating thin shell, an impulsively accelerated circular disk, and a flapping insect wing.

\begin{figure}[ht]                            
   \centering
   \begin{subfigure}{\textwidth}
         \begin{tikzpicture}
            \filldraw[black] (-1,2) node[anchor=north] {$a)$};
            \draw[black, thick, -] (0,-0.5)--(10,-0.5);
            \draw[black, thick, -] (0, 0.5)--(10, 0.5);
            \foreach \x in {0,5,10}{
                \draw[dashed, gray, thick, -] (\x,-1)--(\x,1);
            }
            \filldraw[gray] (0, -1.25) node[anchor=north] {$x=-\frac{L}{2}$};
            \filldraw[gray] (5, -1.25) node[anchor=north] {$x=0$};
            \filldraw[gray] (10, -1.25) node[anchor=north] {$x=\frac{L}{2}$};
            \draw [fill=gray] (4.0,-0.5) rectangle (4.15,0.5);
            \draw[black, thick, ->] (3.5,0.)--(4.5,0.);
            \filldraw[black] (1.2, 0.2) node[anchor=west] {$u^0 = 0$};
            \filldraw[black] (7.2, 0.2) node[anchor=west] {$u^0 = 0$};
            \filldraw[black] (4.4, -0.05) node[anchor=west] {$v_b^{n+1}$};
            \filldraw[black] (4.075,-0.75) node[anchor=north] {$x_b$};
            \draw [red] plot [smooth] coordinates {(0.0, 0.0) (0.15873015873015872, 0.0) (0.31746031746031744, 0.0) (0.47619047619047616, 0.0) (0.6349206349206349, 0.0) (0.7936507936507936, 0.0) (0.9523809523809523, 0.0) (1.1111111111111112, 0.0) (1.2698412698412698, 0.0) (1.4285714285714284, 0.0) (1.5873015873015872, 0.0) (1.746031746031746, 0.0) (1.9047619047619047, 0.0) (2.0634920634920633, 0.0) (2.2222222222222223, 0.0) (2.380952380952381, 0.0) (2.5396825396825395, 0.0) (2.698412698412698, 0.0) (2.8571428571428568, 0.0) (3.015873015873016, 0.0006215393905388278) (3.1746031746031744, 0.07335455918392209) (3.333333333333333, 0.24999999999999972) (3.492063492063492, 0.48753465413096364) (3.6507936507936507, 0.7281053286765813) (3.8095238095238093, 0.9131193871579972) (3.968253968253968, 0.9975153876827006) (4.1269841269841265, 0.960738105935204) (4.285714285714286, 0.8117449009293669) (4.444444444444445, 0.5868240888334649) (4.603174603174603, 0.3407566748741584) (4.761904761904762, 0.133474064085087) (4.92063492063492, 0.015461356885461297) (5.079365079365079, 0.0) (5.238095238095238, 0.0) (5.396825396825396, 0.0) (5.555555555555555, 0.0) (5.7142857142857135, 0.0) (5.873015873015873, 0.0) (6.031746031746032, 0.0) (6.19047619047619, 0.0) (6.349206349206349, 0.0) (6.507936507936508, 0.0) (6.666666666666666, 0.0) (6.825396825396825, 0.0) (6.984126984126984, 0.0) (7.142857142857142, 0.0) (7.301587301587301, 0.0) (7.46031746031746, 0.0) (7.619047619047619, 0.0) (7.777777777777778, 0.0) (7.936507936507936, 0.0) (8.095238095238095, 0.0) (8.253968253968253, 0.0) (8.412698412698413, 0.0) (8.571428571428571, 0.0) (8.73015873015873, 0.0) (8.88888888888889, 0.0) (9.047619047619047, 0.0) (9.206349206349206, 0.0) (9.365079365079364, 0.0) (9.523809523809524, 0.0) (9.682539682539682, 0.0) (9.84126984126984, 0.0) (10.0, 0.0)};
            \draw [blue] plot [smooth] coordinates {(0.0, 0.0) (0.15873015873015872, 0.0) (0.31746031746031744, 0.0) (0.47619047619047616, 0.0) (0.6349206349206349, 0.0) (0.7936507936507936, 0.0) (0.9523809523809523, 0.0) (1.1111111111111112, 0.0) (1.2698412698412698, 0.0) (1.4285714285714284, 0.0) (1.5873015873015872, 0.0) (1.746031746031746, 0.0) (1.9047619047619047, 0.0) (2.0634920634920633, 0.0) (2.2222222222222223, 0.0) (2.380952380952381, 0.0) (2.5396825396825395, 0.0) (2.698412698412698, 0.0) (2.8571428571428568, 0.0) (3.015873015873016, 0.001957849080197308) (3.1746031746031744, 0.2291090123491573) (3.333333333333333, 0.5564331385706446) (3.492063492063492, 0.7482341605125363) (3.6507936507936507, 0.7577976248186956) (3.8095238095238093, 0.5827942842164603) (3.968253968253968, 0.2658474016528158) (4.1269841269841265, -0.11584843750461435) (4.285714285714286, -0.469328595768387) (4.444444444444445, -0.7085005581020912) (4.603174603174603, -0.7751123539718157) (4.761904761904762, -0.6529402239855749) (4.92063492063492, -0.371740027678821) (5.079365079365079, -0.04870327418920309) (5.238095238095238, 0.0) (5.396825396825396, 0.0) (5.555555555555555, 0.0) (5.7142857142857135, 0.0) (5.873015873015873, 0.0) (6.031746031746032, 0.0) (6.19047619047619, 0.0) (6.349206349206349, 0.0) (6.507936507936508, 0.0) (6.666666666666666, 0.0) (6.825396825396825, 0.0) (6.984126984126984, 0.0) (7.142857142857142, 0.0) (7.301587301587301, 0.0) (7.46031746031746, 0.0) (7.619047619047619, 0.0) (7.777777777777778, 0.0) (7.936507936507936, 0.0) (8.095238095238095, 0.0) (8.253968253968253, 0.0) (8.412698412698413, 0.0) (8.571428571428571, 0.0) (8.73015873015873, 0.0) (8.88888888888889, 0.0) (9.047619047619047, 0.0) (9.206349206349206, 0.0) (9.365079365079364, 0.0) (9.523809523809524, 0.0) (9.682539682539682, 0.0) (9.84126984126984, 0.0) (10.0, 0.0)};
            \filldraw[red] (4.2,1.2) node[anchor=west] {$f^{n+1}=\delta(x-x_b)$};
            \filldraw[blue] (3.4,1.2) node[anchor=east] {$\Delta \varphi=\delta'(x-x_b)$};
       \end{tikzpicture}
     \end{subfigure} %
     \hspace{1cm}
     \begin{subfigure}{\textwidth}
          \begin{tikzpicture}
            \filldraw[black] (-1,2) node[anchor=north] {$b)$};
            \draw[black, thick, -] (0,-0.5)--(10,-0.5);
            \draw[black, thick, -] (0, 0.5)--(10, 0.5);
            \foreach \x in {0,5,10}{
                \draw[dashed, gray, thick, -] (\x,-1)--(\x,1);
            }
            \filldraw[gray] (0, -1.25) node[anchor=north] {$x=-\frac{L}{2}$};
            \filldraw[gray] (5, -1.25) node[anchor=north] {$x=0$};
            \filldraw[gray] (10, -1.25) node[anchor=north] {$x=\frac{L}{2}$};
            \draw [fill=gray] (4.0,-0.5) rectangle (4.15,0.5);
            \draw[black, thick, ->] (3.5,0.)--(4.5,0.);
            \filldraw[black] (4.4, -0.05) node[anchor=west] {$v_b^{n+1}$};
            \filldraw[black] (4.075,-1) node[anchor=west] {$x_b$};
            \draw [red] plot coordinates {(0.0, -2*0.1171) (4.0,-2*0.49210)};
            \draw [red, dashed] (4.0,-2*0.49210) -- (4.15, 2*0.49210);
            \draw [red] plot coordinates {(4.15, 2*0.49210) (10, -2*0.1015)};
            \filldraw[red] (4.2,1.55) node[anchor=north] {$\varphi(x)$};
            \filldraw[red] (6.2,0.8) node[anchor=west] {$\nabla\varphi(x)=v_b^{n+1}$};
       \end{tikzpicture}
     \end{subfigure}
     \hspace{1cm}
     \begin{subfigure}{\textwidth}
          \begin{tikzpicture}
            \filldraw[black] (-1,2) node[anchor=north] {$c)$};
            \draw[black, thick, -] (0,-0.5)--(10,-0.5);
            \draw[black, thick, -] (0, 0.5)--(10, 0.5);
            \draw[blue, thick, ->] (1,-0.05) -- (2,-0.05);
            \filldraw[blue] (1.5, -0.15) node[anchor=south] {$u^{n+1} = v_b^{n+1}$};
            \draw[blue, thick, ->] (7,-0.05) -- (8,-0.05);
            \filldraw[blue] (7.5, -0.15) node[anchor=south] {$u^{n+1} = v_b^{n+1}$};
            \foreach \x in {0,5,10}{
                \draw[dashed, gray, thick, -] (\x,-1)--(\x,1);
            }
            \filldraw[gray] (0, -1.25) node[anchor=north] {$x=-\frac{L}{2}$};
            \filldraw[gray] (5, -1.25) node[anchor=north] {$x=0$};
            \filldraw[gray] (10, -1.25) node[anchor=north] {$x=\frac{L}{2}$};
            \draw [fill=gray, fill opacity=0.15] (4.0,-0.5) rectangle (4.15,0.5);
            \draw[gray, thick, ->, opacity=0.5] (3.55,0.)--(4.45,0.);
            \draw [fill=gray] (4.925,-0.5) rectangle (5.075,0.5);
            \draw[black, thick, ->] (4.55,0.)--(5.45,0.);
            \filldraw[black] (5.4, -0.05) node[anchor=west] {$v_b^{n+1}$};
            \filldraw[black] (4.8,-1) node[anchor=west] {$x_b$};
            \draw [orange, thick, ->] plot [smooth] coordinates {(4.075, 0.55) (4.55, 0.8) (5, .55)};
            \filldraw[orange] (4.55, 0.8) node[anchor=south] {$t = t_0+\Delta t$};
       \end{tikzpicture}
     \end{subfigure}
   \caption{Schematic of the 1D periodic pipe with piston problem. The problem is periodic such that any quantity $\theta$ respects $\theta(x=-L/2)=\theta(x=L/2)$. The thin wall piston is initially located at $x=x_b$. The immersed forces $f^{n+1}$ is added to the initial quiescent fluid field in $a$). This immersed forcing generated a dipole source term for the pressure Poisson equation $a$). The resulting pressure field after successful inversion of the Poisson equation with the correct $f_p$ in $b$). The final velocity field with the piston in its new position in $c$).}
   \label{fig:pipe}
\end{figure}
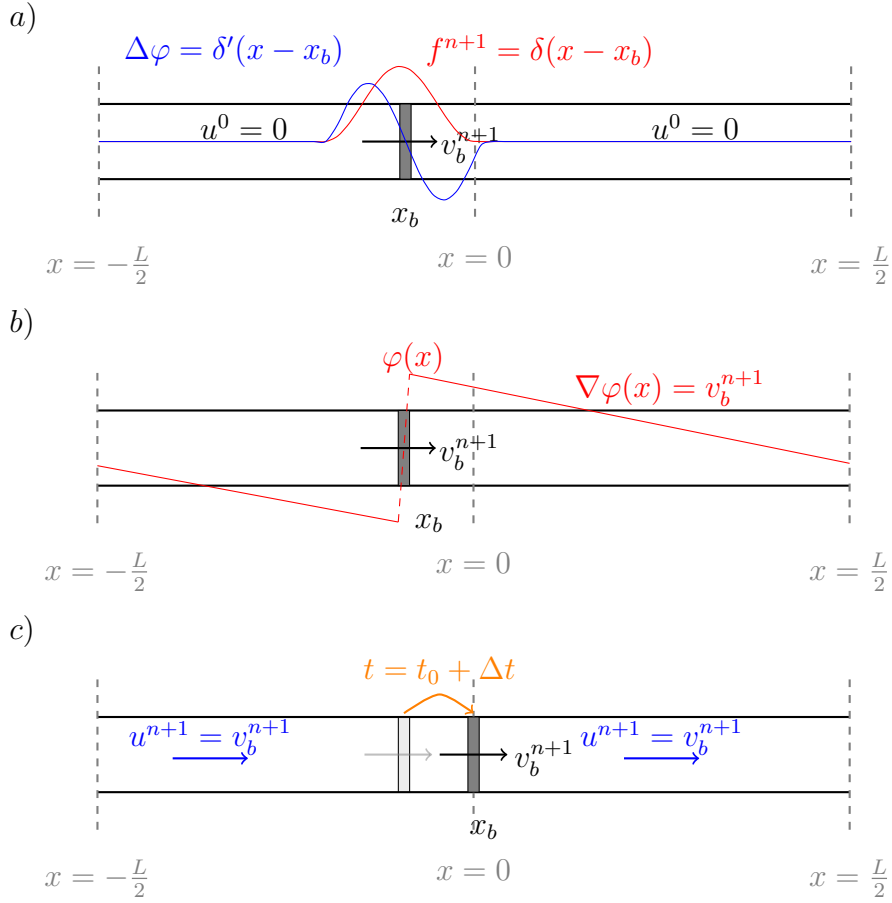

\section{Illustrative one-dimensional model}
\label{S:1D}

We start by demonstrating the errors introduced by incorrect pressure treatment in immersed boundary methods using a simple unsteady 1D-piston problem with an analytical solution. This is a simplification of the full 3D flow, which ignores the cross-section variation to focus on the driving impulse of the piston on the fluid. This simple system will allow us to directly compare the performance of a variety of immersed boundary methods.


\subsection{Problem definition}

We consider the flow inside a periodic 1D domain, in which a thin-wall piston is immersed, Figure~\ref{fig:pipe}. The flow and the piston inside the domain are initially quiescent, and the piston body has unsteady velocity $v_b(t)$ where $v_b(0)=0$. The flow is assumed to be incompressible and viscous effects with the wall are omitted. The momentum equation in the direction of motion $x$ is then simply
\begin{equation}\label{eq:1}
    \frac{\partial u}{\partial t} = -\frac{1}{\rho}\frac{\partial p}{\partial x},
\end{equation}
where $u,p$ are the 1D velocity and pressure fields and $\rho$ is the constant fluid density. Velocity continuity is simply
\begin{equation}\label{eq:2}
    \frac{\partial u}{\partial x} = 0.\\
\end{equation}
This is complemented with adequate boundary conditions on the velocity and pressure fields at the piston location $x_b$
\textcolor{black}{
\begin{subequations}
\begin{align}
    u(x_b,t) = v_b(t), \label{vbc} \\
    \frac{\diff p}{\diff x}\Big\rvert_{x_{b}} = -\rho\frac{\diff v_b}{\diff t}. \label{pbc}
\end{align}
\end{subequations}
}
The solution at time $t$ must be $u(x,t) = v_b(t)$ in order to satisfy continuity (\ref{eq:2}) and the velocity boundary condition (\ref{vbc}).


\subsection{Immersed boundary formulation and solution}
\textcolor{black}{We demonstrate the necessity to correctly impose (\ref{pbc}) to solve this simple problem. Immersed boundary algorithms generally re-write equation~(\ref{eq:1},~\ref{eq:2}) in the form of a momentum update and pressure Poisson equation to obtain the solution at $t^{n+1}=t^{n}+\Delta t$ from the solution at $t^n$
\begin{subequations}
\begin{align}
    &u^{n+1} = u^n + r_{\Delta t}(u^n)- \frac{\Delta t}{\rho}\nabla p^{n+1} + f^{n+1} \label{eq:PJM1}\\
    &\nabla\cdot \left(\frac{\Delta t}{\rho}\nabla p^{n+1}\right) = \nabla\cdot \left(u^n + r_{\Delta t}(u^n) + f^{n+1}\right)\label{eq:PJM2}
\end{align}
\end{subequations}
where the term $r_{\Delta t}$ contains all the convective and viscous terms (which do not appear in this 1D example). The time level of the pressure term has been discussed in \citep{BlairPerot1993,Strikwerda1999,Brown2001}.
\\
\indent The forcing $f^{n+1}$ in equations (\ref{eq:PJM1},\ref{eq:PJM2}) enforces the velocity and pressure boundary conditions (\ref{vbc},\ref{pbc}) from time $t^n$ to $t^{n+1}$. In general, this forcing term can be a function of both the body velocity $v_b$ and the fluid state $u,p$. This forcing is constrained to be local to the body, the piston position $x_b$ in this case, using a Dirac delta function
\begin{equation}\label{eq:forcing}
    f^{n+1} = \delta (x-x_b)C(v_b,u,p)
\end{equation}
for some function $C(v_b,u,p)$ that imposes the boundary conditions, see Figure~\ref{fig:pipe}a. Since the fluid is initially still and letting $\varphi=\Delta t p^{n+1}/ \rho$, we seek a solution to
\begin{equation}
    \nabla^2\varphi = \nabla\cdot f^{n+1} = \nabla\cdot\delta(x-x_b)C(v_b,u,p).
\end{equation}
Because the Poisson equation is linear, we can split the function $C$ in two functions, one for the velocity boundary condition $f_u = f_u(v_b,u)$ and one for the pressure boundary condition $f_p = f_p(v_b,p)$.
\begin{equation}
    \nabla^2\varphi = \delta'(x-x_b)f_u + \delta'(x-x_b)f_p + \delta(x-x_b)f_p',
\end{equation}
where the last term appears due to the spatial dependency of the pressure boundary condition. Using the definition of the Dirac delta function as the derivative of the Heaviside function $\nabla H(x-x_b)=\delta(x-x_b)$, the solution to (\theequation) is
\begin{equation}
    \varphi = H(x-x_b)(f_u+f_p) + \nabla^{-2}\delta(x-x_b)f_p',
\end{equation}
where $\nabla^{-2}$ is the inverse Laplacian. This solution is illustrated in Figure~\ref{fig:pipe}b, which shows the jump in the pressure across the piston body due to the Heaviside function, while the pressure boundary condition imposes the constant pressure gradient in the rest of the periodic domain.
\\
\indent The velocity field is simply
\begin{equation}
    u^{n+1} = f^{n+1} - \nabla\varphi = -\nabla\nabla^{-2}\delta(x-x_b)f_p',
\end{equation}
where the $f^{n+1}$ terms cancel, revealing that the solution depends \textit{only} on the treatment of the pressure boundary condition for this canonical problem. Note that the final velocity field satisfies condition (\ref{vbc}) as long as the pressure condition (\ref{pbc}) is met, since substituting (\ref{pbc}) into the time integrated governing equation (\ref{eq:1}) gives (\ref{vbc}).}

\subsection{Forcing terms with and without a pressure boundary condition}

\begin{table}
\centering
    \caption{Classification of the different immersed boundary methods and associated error on the 1D problem for zero thickness body. Error is quantified as $L_\infty(u^{n+1}/v_b^{n+1}-1)$.}
    \begin{tabular}{lll}
    \hline
    Method &  Reference & Error\\
    \hline
    IB  &  \cite{Peskin72} & $0.9375$\\
    cIBM &   \cite{Bale2020} & $0.9717$\\
    Direct-forcing  &  \cite{Fadlun2000} & $0.9375$\\
    BDIM &  \cite{Maertens2015}& $0.7239$\\
    IBPM &  \cite{Colonius2008}& $1.483\times 10^{-11}$\\
    BDIM-$\sigma$ & - & $3.306\times 10^{-10}$ \\
    \hline
    \end{tabular}
    \label{tab:EBM_description}
\end{table}

\textcolor{black}{Table~\ref{tab:EBM_description} summarizes six immersed boundary approaches and their accuracy when applied to this canonical test case (found online at \cite{github}). The previous section showed that the forcing $f^{n+1}$ must explicitly contain a term that deals with the pressure boundary condition for problems which are pressure-driven, such as this piston flow. As such, the methods which do not include $p^{n+1}$ in $f^{n+1}$ (IB, cIBM, and Direct-forcing) show approximately 100\% error in their solution. 
\\
\indent BDIM and the new BDIM-$\sigma$ approach presented in this manuscript use an update of the form
\begin{equation}
    f^{n+1} =  (1-\mu^0)\left[ v_b^{n+1} - u^n - r_{\Delta t}(u^n) +\frac{\Delta t}{\rho}\nabla p^{n+1} \right]
\end{equation}
where $1-\mu^0 \sim \delta(x-x_b)$ is a kernel function centered on the body. The first term imposes (\ref{vbc}) on the body, while the second and third terms adjust the convection/diffusion and pressure forcing applied to the fluid close to the body. With the piston initially at rest, the pressure equation becomes
\begin{equation}
    \nabla \cdot \frac{\mu^0 \Delta t}{\rho}\nabla p^{n+1} = \nabla\cdot (1-\mu^0)v_b^{n+1}.
\end{equation}
where we can see that the addition of the pressure term in $f^{n+1}$ has resulted in a Poisson equation with variable coefficients. If these coefficients are properly developed, they enforce the pressure boundary condition and allow a jump in pressure across the interface, leading to the correct solution. As detailed in the next section, this requires a modification to BDIM for very thin geometries, which we call BDIM-$\sigma$. The error is reduced to machine precision using this approach. 
\\
\indent The IBPM method also obtains good results on this problem but it requires a coupled momentum/pressure equation
\begin{equation}
    \nabla^2 \lambda = \nabla\cdot r_1,
\end{equation}
where $\lambda \equiv [\frac{\Delta t}{\rho}p^{n+1},f_u]^\top$ and $r_1 = [0,v_b^{n+1}]^\top$, such that the pressure and the velocity forcing are determined simultaneously to enforce both the no-slip and the divergence-free constraint. The primary methodological contribution of this manuscript is that BDIM-$\sigma$ achieves this high level of accuracy while using a standard Poisson equation.}

\section{A minimal thickness Boundary Data Immersion Method}
\label{bdimsigma}

The previous section illustrates that the projection step must explicitly impose the boundary condition on the pressure for the results to be accurate. This section generalizes that result for immersed surface simulations of the two and three dimensional Navier-Stokes equation. In particular, we demonstrate the requirement of a finite body thickness for the Boundary Data Immersion Method (BDIM) and develop a new approach to simulate the flow past thin surfaces.

As detailed in \cite{Maertens2015} BDIM solves a \textit{meta}-equation for the velocity field $\vec u$ developed from a convolution of the Navier-Stokes equations in the fluid domain $\Omega_f$ with the prescribed velocity in the body domain $\Omega_b$. First we write the fluid equation in the form of an update over a time step $\Delta t=t^{n+1}-t^n$
\begin{subequations}\label{eq:NSupdate}
    \begin{align}
     \vec u(t^{n+1},\vec x) &= \vec u(t^n,\vec x)+\vec r_{\Delta t}(\vec u(t,\vec x))-\vec \nabla p_{\Delta t}(\vec x), \quad \forall \vec x \in \Omega_f, \\
     \vec r(\vec u) &= \frac 1{Re} \nabla^2\vec u - \vec u \cdot \vec\nabla\vec u,
     \end{align}
\end{subequations}
where we use the notation $f_{\Delta t}=\int_{t^n}^{t^{n+1}}f\diff{t}$ for the impulse of $f$ across the time step, $\vec \nabla p_{\Delta t}$ is the pressure impulse \textcolor{black}{ with $p\equiv P/\rho U^2$ the non-dimensionnal pressure ($U$ is a velocity length-scale)}, $\vec r_{\Delta t}$ is the impulse of all non-pressure terms and $Re$ is the Reynolds number. The body velocity update is simply:
\begin{equation}\label{eq:vel_bc}
    \vec u(t^{n+1},\vec x) = \vec v_b(t^{n+1},\vec x),\quad \forall \vec x \in \Omega_b.
\end{equation}
We can extend the domain of application of the fluid and the solid governing equations onto the full domain $\vec x \in \Omega=\Omega_f\cup\Omega_b$ by convolution of the equations
using a kernel with support $\epsilon$
\begin{equation}\label{eq:kernel}
    \phi_\epsilon(|x-x'|) = \begin{cases} \frac{1}{2\epsilon}\left(1+\cos\left(\frac{|x-x'|\pi}{\epsilon}\right)\right) & \text{ if } |x-x'|<\epsilon\\
    0 & \text{else}.
    \end{cases}
\end{equation}
Evaluation of this convolution using a Taylor series expansion up to second order and assuming that the boundary can be approximated by its local tangential plane we obtain an equation for the convolved velocity field $\vec u_\epsilon$ valid throughout the domain
\begin{equation}\label{eq:BDIMupdate}
\begin{split}
    \vec u_\epsilon^{n+1} = \mu^0\,\left(\vec u_\epsilon^n +\vec r_{\Delta t}(\vec u_\epsilon^n)-\vec\nabla p_{\Delta t} \right) + (1-\mu^0)\vec{v}_b\\
    +\mu^1\frac{\partial}{\partial n}\left(\vec u_\epsilon^n+\vec r_{\Delta t}(\vec u_\epsilon^n)-\vec v_b-\vec\nabla p_{\Delta t}\right),
\end{split}
\end{equation}
where we define $\mu^k(s)=\int_{-\epsilon}^s \phi_\epsilon(x)x^k \diff{x}$ as the moments of the kernel and  $s(\vec x,t)$ is the signed distance function to the boundary $\partial\Omega_b$. See \cite{Maertens2015} for a full derivation. The key properties of this meta-equation are (i) the kernel moments \textit{smoothly} ramp between the fluid and solid governing equations, and (ii) this transition is applied not only to the velocity and convection/diffusion impulse ($u_\epsilon,v_b,r_{\Delta t}$) but \textit{also} to the pressure impulse.

Equation~\ref{eq:BDIMupdate} is solved using the fractional step method, where the pressure is determined by requiring that the final velocity field be divergence-free. Heun's method is use to integrate the equation \textcolor{black}{ form $\vec u^n = \vec u_\varepsilon(t^n)$ to find $\vec u^{n+1} = \vec u_\varepsilon(t^{n+1})$ in two steps, first an explicit Euler step is performed using $\vec v_b = \vec v_b(t^{n+1})$
\begin{subequations}
\begin{align}
    &\vec u^* = \mu^0\,\left(\vec u^n+\vec r_{\Delta t}(\vec u^n)\right) + (1-\mu^0)\vec{v}_b + \mu^1\frac{\partial}{\partial n}\left(\vec u_\epsilon^n+\vec r_{\Delta t}(\vec u^n)-\vec v_b\right),\label{eq:Euler1} \\
    &\Delta t\vec\nabla\cdot\left(\mu^0\vec \nabla  p_0\right) = \vec\nabla\cdot \vec u^*, \label{eq:Euler2}\\
    & \vec u_{1} = \vec u^*-\Delta t\mu^0\vec\nabla p_0,\label{eq:Euler3}
\end{align}
\end{subequations}
followed by Heun's corrector step
\begin{subequations}
    \begin{align}
    &\vec u_1^* = \mu^0\,\left(\vec u^n +\vec r_{\Delta t}(\vec u_1)\right) + (1-\mu^0)\vec{v}_b + \mu^1\frac{\partial}{\partial n}\left(\vec u_0+\vec r_{\Delta t}(\vec u_1)-\vec v_b\right), \label{eq:Euler_21}\\
    &\Delta t\vec\nabla\cdot\left(\mu^0\vec \nabla  p_2\right) = \vec\nabla\cdot \vec u_1^*, \label{eq:Euler_22}\\
    & \vec u_2 = \vec u_1^*-\Delta t\mu^0\vec\nabla p_2,\label{eq:Euler_23}\\
    &\vec u^{n+1} = \frac{1}{2} (\vec u_1 + \vec u_2),\label{eq:Euler_24}
    \end{align}
\end{subequations}}
where all the impulses are approximated using explicit methods.

\subsection{Pressure boundary condition enforcement}

\textcolor{black}{At a non-slip wall, the pressure can be shown to satisfy the condition
\begin{equation}\label{eq:fullpbc}
    \frac{\partial p}{\partial n} = \hat{n}\cdot\left[-\rho\frac{\text{D} \vec v_b}{\text{D}t} + \mu\nabla^2\vec u\right].
\end{equation}
where $\text{D}\vec v_b/\text{D}t$ is the acceleration of the moving wall. In this section we will show that this condition can only be imposed by explicit modification of the coefficients of the Poisson equation. We will use our 1D case to show that in the limit of vanishing kernel width, our method imposes this boundary condition on the pressure field implicitly. This argument extends to higher dimensions.}

The pressure is determined through the solution of the Poisson equation in \ref{eq:Euler2}. As opposed to other immersed methods, BDIM uses a \textit{variable} coefficient Poisson equation due to the kernel moment $\mu^0$. A distance $\pm\epsilon$ from the surface, the kernel moment ranges from  $\mu^0(s\ge\epsilon)=1$ outside the body to $\mu^0(s\le-\epsilon)=0$ within the body. The influence of the pressure gradient is removed when $\mu^0=0$, turning off the pressure sensitivity across the body and enforcing the Neumman pressure condition.

We demonstrate this using the 1D example of the previous section since the argument generalizes directly to the 2D and 3D cases. On a 1D uniform grid, a finite volume integration over cell $\Delta\Omega_i$ and central difference of a derivative at cell face $i+\frac12$ gives
\begin{equation}
  \int_{i-\frac 12}^{i+\frac 12} \frac{d\phi}{dx} \diff{x} = \phi_{i+\frac 12}-\phi_{i-\frac 12}\quad \text{and}\quad \frac{dp}{dx}\Big\rvert_{i+\frac 12} = \frac{p_{i+1}-p_i}{\Delta x},
\end{equation}
where $\Delta x$ is the uniform spacing. Substitution into Equation~\ref{eq:Euler2} gives the discrete Poisson equation
\begin{equation}
  \mu^0_{i+\frac{1}{2}}\left(p_{i+1}-p_{i}\right) -\mu^0_{i-\frac{1}{2}}\left(p_i-p_{i-1}\right)
  =\frac{\Delta x}{\Delta t} \left(u^*_{i+\frac{1}{2}}-u^*_{i-\frac{1}{2}}\right).
\end{equation}

Now, consider when the boundary $\partial\Omega_b$ is located at $i+\frac 12$. Letting the support of the kernel $\epsilon\to0$ such that $\mu^0_{i+\frac{1}{2}}=\mu^1_{i\pm\frac{1}{2}}=0$ while $\mu^0_{i-\frac 12}=1$ gives a pressure Poisson equation
\textcolor{black}{\begin{equation}
  -\left(p_{i}-p_{i-1}\right) =\frac{\Delta x}{\Delta t}\left(v_b^{n+1}-u^*_{i-\frac{1}{2}}\right),
\end{equation}}
where $u^*_{i+\frac{1}{2}}=v_b^{n+1}$ from Equation~\ref{eq:Euler1}. This is \textit{exactly} the Poisson equation we would obtain by substituting the velocity and pressure boundary condition (\ref{vbc}),(\ref{pbc}) into a constant coefficient ($\mu^0=1$ everywhere) Poisson equation at this boundary point. \textcolor{black}{This is a 1D discretization of (\ref{eq:fullpbc}) where the viscous term has vanished due to the unidimensionality of the problem}. This same argument applies to 2D and 3D Poisson equations as well; the pressure condition (\ref{eq:fullpbc}) is automatically applied to all immersed surfaces as $\mu^0\to0$ in the body.
\textcolor{black}{The zeros in the Poisson matrix ensure that condition (\ref{eq:fullpbc}) is imposed as we approach the body and that pressure jumps across the interface do not influence the fluid momentum on either side, which would result in an leakage of fluid across the body. The resulting} linear system can still be solved with standard iterative methods (relaxation, Krylov subspace, Multi-Grid, etc), unlike IBPM which requires careful construction of the augmented system to ensure that it is positive-definite and well-conditioned \cite{Taira2007_1}.

\subsection{Acceleration and Reynolds number dependence}

We note that the error introduced by the lack of explicit boundary condition on the pressure field is dependent on the strength of the r.h.s of equation~\ref{eq:Euler2}. The major contributor to the magnitude of this source term in the Poisson equation is the local body acceleration (through the $v_b^{n+1}$ term). However, the Reynolds number of the flow will also influence the strength of this source term. Low Reynolds number flow is characterized by a relatively smooth velocity gradient near walls compared to high Reynolds number flow, therefore reducing the contribution of $r_{\Delta t}$ to the source term. This means that the errors in immersed boundary methods which do not enforce the Neumann condition in the project step will be more severe as the Reynolds number increases, which may explain why this issue is not referenced more commonly in the literature of low $Re$ immersed boundary simulations. We will demonstrate the magnitude and increasing severity of these errors with $Re$ in Section~\ref{S:Disk}.

\subsection{BDIM-$\sigma$}
\label{sec:bdim-sigma}

\begin{figure}
    \centering
    \includegraphics[scale=0.65]{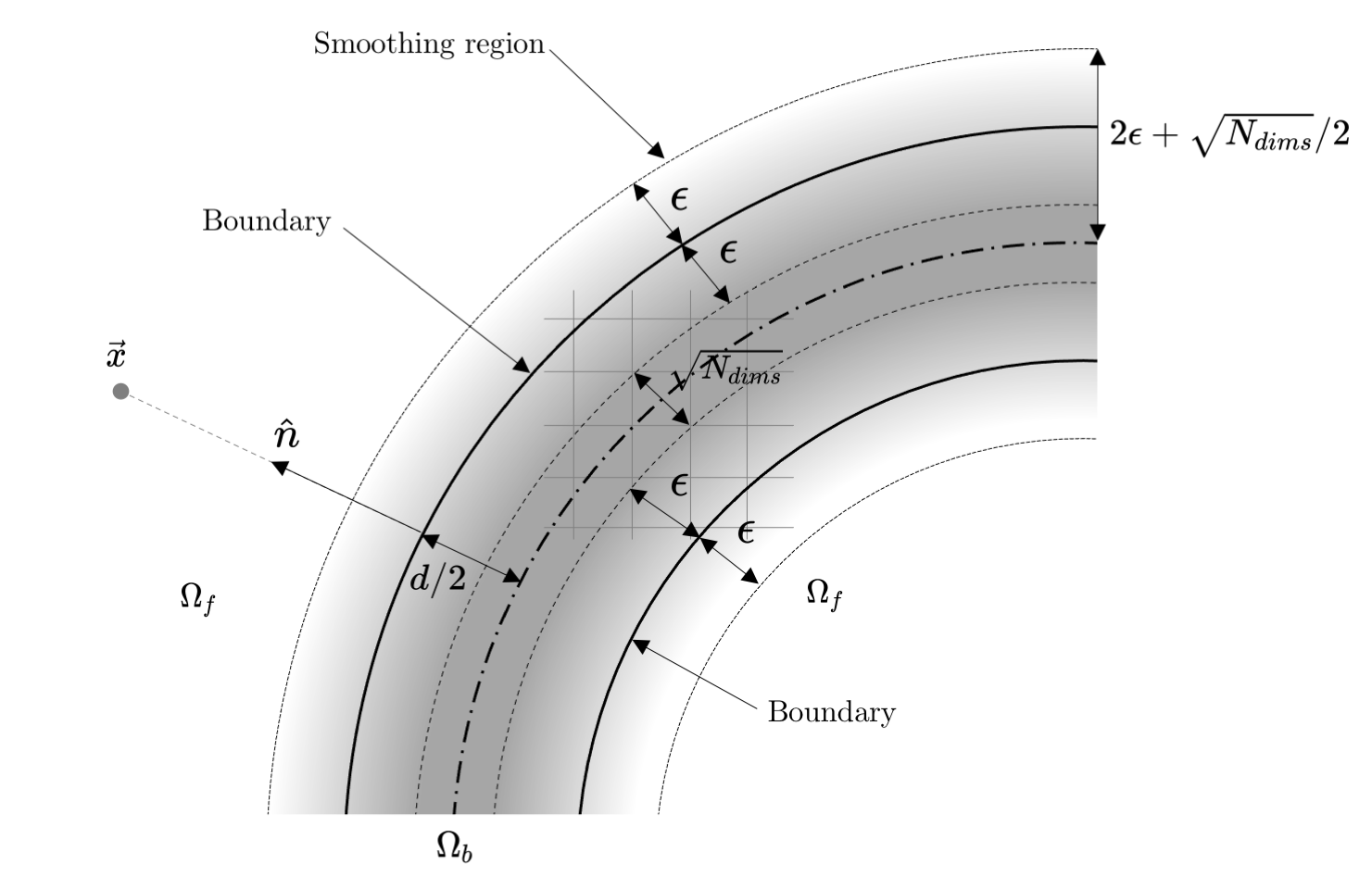}
    \caption{Schematic of the BDIM smoothing for a general thin 2D body. The smoothing occurs in a region $\epsilon$ away from the boundary, represented here by the gray gradient area. A schematic of a 2D grid is also depicted to show the minimum width of the core of the body (where $\mu^0=0$) that ensure the Neumann condition is respected.}
    \label{fig:kernel_lims}
\end{figure}

We have shown that the Neumann condition is imposed for boundary cells if $\mu^0=0$ within the body. With a kernel support of $\epsilon = 2\Delta x$, as suggested in \citep{Maertens2015}, the smoothing region on each side of a 1D body has a width of $4\Delta x$. If the center of the body is aligned with a grid point, we would satisfy $\mu^0=0$ for this point. In general, however, the body will not be aligned with a grid point and we must increase the width of the central region of the $\mu^0$ kernel moment to ensure that this is true regardless of the position of the body. Therefore, in 1D we must increase the body thickness to $5\Delta x$ to satisfy the requirement and in $N_{dims}$ we find that the minimum body thickness is $(4+\sqrt{N_{dims}})\Delta x$, Figure~\ref{fig:kernel_lims}. This means that accurate predictions around very thin bodies in 3D space requires decreasing $\Delta x$ at the cost of significantly more expensive simulations. 

In this section, we present BDIM-$\sigma$, a simple adjustment to the classic BDIM method which enables the accurate simulation of thin bodies and even $\mathbb{R}^2$ bodies (plates, shells, membranes) immersed in a $\mathbb{R}^3$ fluid domain. In particular, we follow the common \textit{slender body} formulation that parametrizes a thin 3D body by a mid-plane surface $\sigma$ and a local thickness $d$, see \cite{Chen2014,Duong2017}. In other words, the points on the body surface are defined as
\begin{equation}
    \vec x \in \partial\Omega_b = \vec\sigma(\vec\xi ) \pm \tfrac 12 d(\vec\xi ) \hat n(\vec\xi ) 
\end{equation}
where $\vec\xi$ are the curvilinear coordinates of the mid-surface and $\hat n$ is the unit normal.

Our aim is to produce a zeroth kernel moment $\mu^0$ which behaves similarly to a top-hat function; rapidly but smoothly transitioning from 1 to 0 to 1 as we travel through an immersed geometry of any thickness, Figure~\ref{fig:kernel_lims}. To achieve this, we first minimize the size of the smoothing region by setting $\epsilon=\Delta x/2$. This makes the meta-function transition more abrupt, but the following sections demonstrate that the results are still numerically stable and accurate. Second, we adjust the signed distance function to ensure that the flat $\mu^0=0$ section is always present, regardless of the body thickness $d$
\begin{equation}
    s(\vec x, t) = s_\sigma(\vec x, t) - \max{(\tfrac 12 d(\vec x , t),\ C_1\Delta x)},
\end{equation}
where $s_\sigma$ the signed distance to the surface $\sigma$, $d(\vec x , t)$ is the local thickness at the normal projection point of $\vec x$, see Figure~\ref{fig:kernel_lims}, and $C_1=\frac 12+\frac 12\sqrt{N_\text{dims}}$ is the minimum thickness coefficient which depends on the number of spatial dimensions $N_\text{dims}$. This new adjusted signed distance function ensures the proper boundary condition will be applied, even in the limit of a vanishing thickness, i.e. a $\mathbb{R}^2$ surface in $\mathbb{R}^3$ space. \textcolor{black}{As $d\rightarrow0$, the $\mu^0=0$ section extends outside the body the minimum amount required to maintain the modified Poisson equation.}

\section{Convergence Study: Oscillating Thin Shell}
\label{S:Convergence}

\begin{figure}
    \centering
    \includegraphics[scale=1]{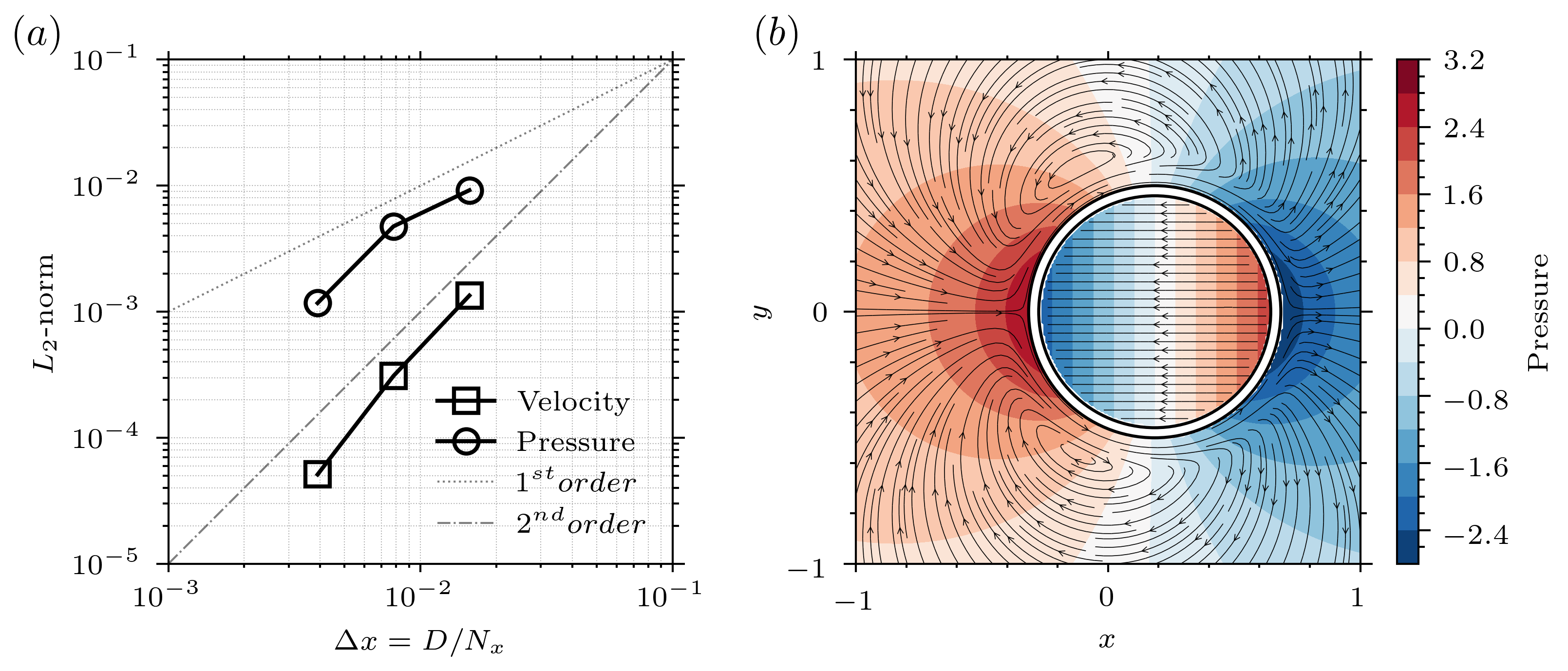}
    \caption{$(a)$ Convergence in the $L_2$-norm of the pressure and velocity field. $(b)$ pressure contour and velocity streamlines for the external and internal flow at $t=0.5$. The internal flow follows rigid body motion.}
\label{fig:spherical_shell}
\end{figure}

\textcolor{black}{In this section we verify the numerical convergence of our implementation of BDIM-$\sigma$ using the 2D test case of an oscillating circular shell in quiescent fluid, equivalent to the test used in \cite{Fadlun2000} and presented in \cite{Gilmanov2005}. A thin shell of diameter $D$ and thickness $d=D/16$ is placed in the center of a box of size $H=8D$ filled with incompressible viscous fluid at rest. The inside of the shell also contains incompressible viscous fluid at rest, Figure~\ref{fig:spherical_shell}b. Fluid flow is obtained by prescribed motion of the shell
\begin{equation}
    x(t) = x_0 + A\left(1-\cos(2\pi ft)\right), \quad y(t) = y_0,
\end{equation}
where the amplitude of motion is $A=0.125D$, the frequency $f=1$ $Hz$, and $\vec x_0=\vec x(t=0)$ is the initial position of the shell. The Reynolds Number of the flow is $Re=UD/\nu=200$, where the reference velocity is $U=2\pi f A$.
\\
\indent The simulations are performed using an in-house code, Lotus, that has been validated for a wide range of flows at intermediate Reynolds numbers, with and without immersed bodies~\cite{Maertens2015,Garcia2019,Fernando2020,zurman_2020}. The governing equations are solved on a Cartesian finite-volume mesh. The convective term is approximated using a flux-limited Quadratic Upstream Interpolation for Convective Kinematics (QUICK) \citep{LEONARD1979} scheme for stability, and central difference is used for the diffusive terms. Turbulence is modelled using implicit Large Eddy Simulation (iLES) model that uses flux limiting to model the energy loss due to sub-grid stress. The solution to the Poisson equation are obtained using a multi-grid method. We use adaptive time-steps based on the convective and diffusive Courant velocities. The second-order spatial and temporal convergence of the flow solver without immersed boundaries is demonstrated on the 2D Taylor-Green vortex in appendix~\ref{A:1}.}

\textcolor{black}{Simulations for the 2D oscillating shell problem are produce using four uniform meshes of size $D/\Delta x \in [32, 64, 128, 256]$. Figure~\ref{fig:spherical_shell}b shows a snapshot of the solution. The results show the expected internal and external flow pattern and pressure field, including the discontinuous jump in pressure across the thin shell wall. Quantitatively, the internal flow is captured perfectly, without any variations between mesh size, as the velocity field solution is constant and the pressure is linear. As the motion amplitude is finite, a closed-form solution for the external flow does not exist. As such, we use the results of the finest mesh to define the $L_2$-norm of the error for a variable $\varphi$}
\begin{equation}\label{eq:L2}
    L_2^N(\varphi) = \left[\frac{1}{N^2}\sum_{i=1}^{N^2}\vert\varphi_i^N - \varphi_i^e\vert^2\right]^{1/2}.
\end{equation}
\textcolor{black}{Figure~\ref{fig:spherical_shell}a shows the second order-convergence (dashed-line) for the pressure ($p=2.0$) and velocity ($p=2.0$) field in the $L_2$-norm, verifying the numerical accuracy of the proposed BDIM-$\sigma$ method.}

\section{Application: Accelerating Thin Disk}
\label{S:Disk}

In this section we validate the performance of the new BDIM-$\sigma$ approach compared to other immersed boundary methods on the pressure-driven flow around an impulsively accelerated thin circular disk.
\begin{figure}[!ht]
    \centering
    \includegraphics[scale=1.]{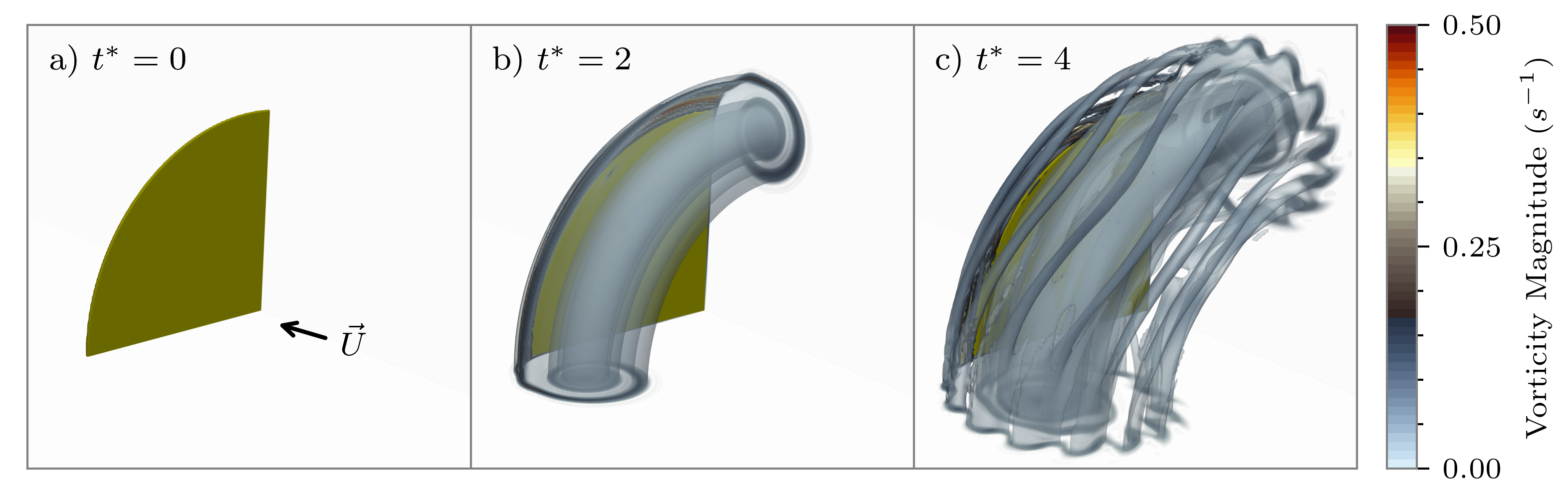}
    \caption{Time snapshot of the disk and the vortex structure visualized using the iso-surface of $\lambda_2 U^2/D^2=-32$, coloured with vorticity magnitude $|\omega| U/D$. Disk and vortex core shown at initial impulse, end of acceleration phase, and in the steady phase. Only $1/4$ of the disk and vortex core are shown. Animation of this figure is provided as supplementary material.}
\label{fig:disk}
\end{figure}
Following the experimental and numerical study presented in \cite{Fernando2020}, an initially static circular disk with diameter $D$ is accelerated from rest with constant acceleration $a$ in a quiescent fluid up to a terminal velocity $U$, after which the velocity is held steady. Figure~\ref{fig:disk} visualizes the resulting flow for non-dimensional acceleration $a^*=aD/U^2=0.5$ at $t^*=tU/D=0$, $t^*=2$ (the end of acceleration), and $t^*=4$. The Reynolds number at the terminal velocity is $Re=UD/\nu=1.25\times10^5$.

\begin{figure}
    \centering\includegraphics[width=\textwidth]{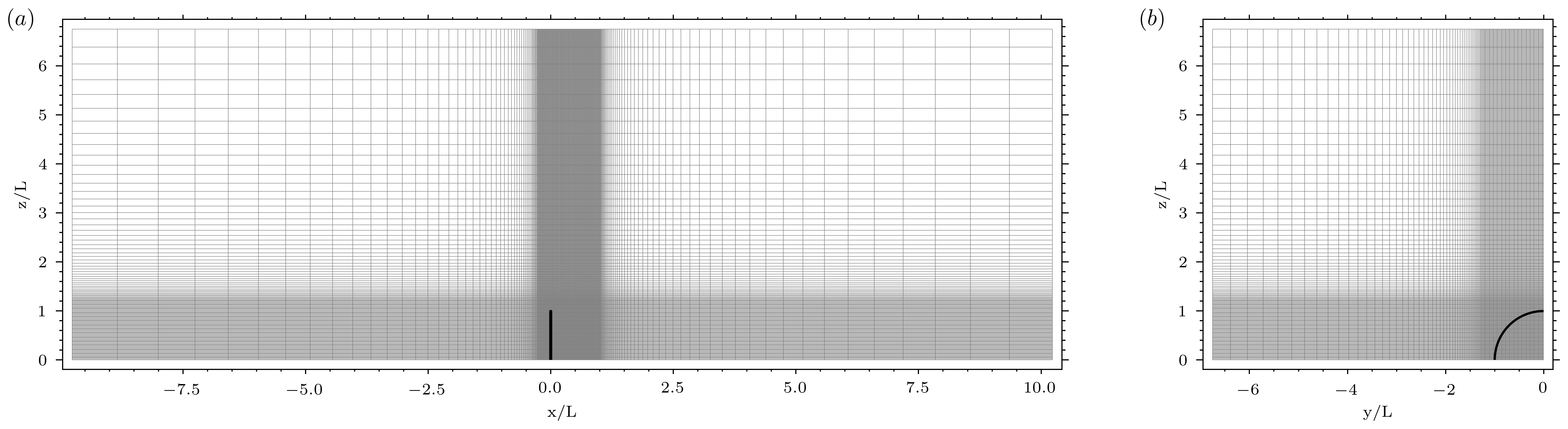}
    \caption{Profile and front view of the finest grid used for the thin circular disk case. The top view is identical to the profile view. For clarity every two cells are shown in each direction.}
    \label{fig:grid}
\end{figure}

We compare three different immersed boundary methods, a Direct-forcing method that uses linear velocity reconstruction \citep{Fadlun2000}, classic BDIM \citep{Maertens2015}, and the new BDIM-$\sigma$ developed above. \textcolor{black}{As we are focused on the early development of the flow, we make use of the double symmetry of the geometry and wake to reduce the mesh count by simulating the flow over $1/4$ of the disk. (The wake breaks symmetry only after $t^*>20$ even at high $Re$ \cite{Fernando2020}.) The no-slip boundary condition is applied on the disk. A zero normal flux condition is applied at the exterior domain boundaries. A uniform grid region of $[1.25\times1.25\times1.25]$ disk radius is used to properly capture the vortex roll-up and the wake of the disk, see Figure~\ref{fig:grid}. Grid stretching is used to fill the rest of the domain until it reaches the total size of $[20\times6.66\times6.66]$ disk radius. Instead of accelerating the disk itself, we accelerate the fluid inside the domain. This ensures that the wake is always located in the uniform region of the grid. Inlet boundary conditions are prescribed as $\vec{u}=[V(t),0,0]$, where $V(t)=at$ for $0<t<U/a$ and $V(t)=U$ for $t\ge U/a$. A convective condition is applied at the domain exit plane downstream of the body.} 

\begin{figure}
    \centering
    \includegraphics[scale=1.]{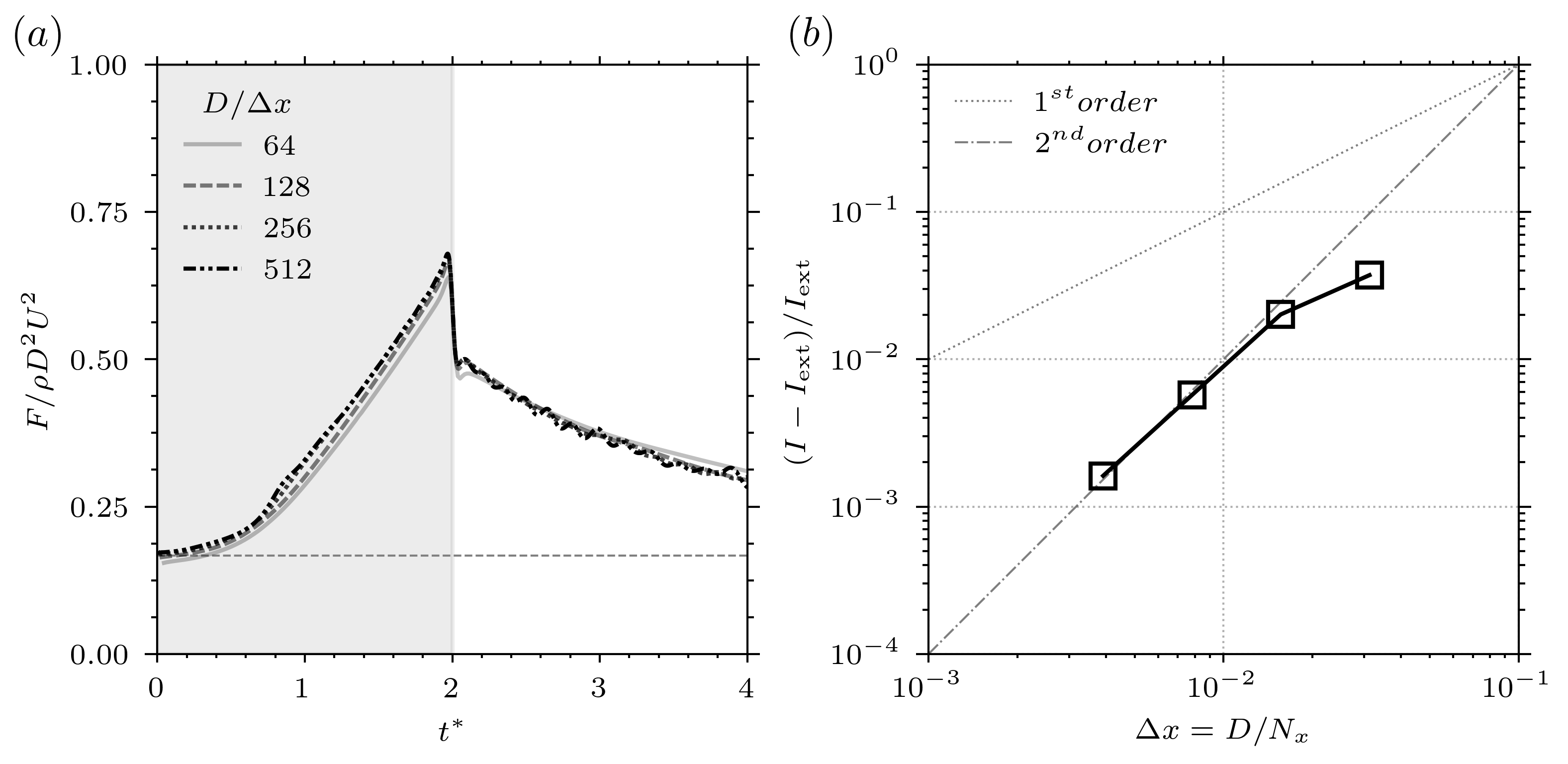}
    \caption{Convergence of the pressure impulse for the thin circular disk with constant thickness ratio $D/d=32$ and varying resolution. $(a)$ Normalized pressure force acting on the disk for varying resolution and $(b)$ relative error in the pressure impulse to the Richardson extrapolated value.}
    \label{fig:conv_disk}
\end{figure}

\textcolor{black}{
We perform two convergence studies for the disk case; in the first one, we only vary the spatial resolution, and in the second one, we also vary the thickness of the body. Convergence is assessed by comparing the pressure impulse $I \equiv \int_0^T F/\rho D^2  \diff t^*$ where we have taken $T=4t^*$ for all cases. Richardson extrapolation is used to obtain the extrapolated value and apparent order of convergence.}

\begin{figure}
    \centering
    \includegraphics[scale=1.]{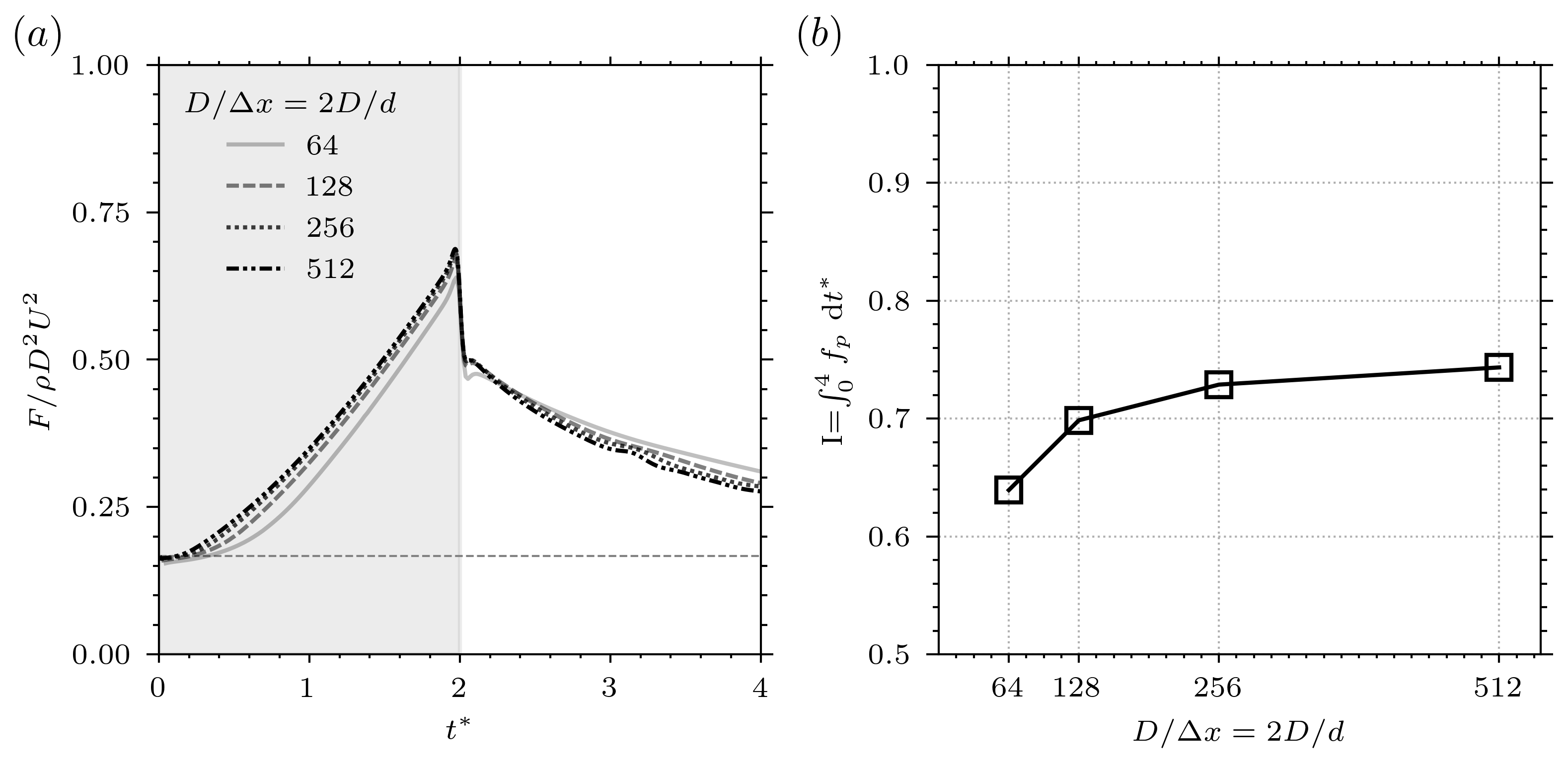}
    \caption{Thickness study for the thin circular disk with the thickness reduced simultaneously with the grid spacing, maintaining a minimal $d=2\Delta x$ at each resolution. $(a)$ normalized pressure force on the disk at various resolution, $(b)$ variation of the pressure impulse for varying resolution and thicknesses.}
    \label{fig:conv_disk_2}
\end{figure}

\textcolor{black}{We obtain the expected $2^{nd}$-order convergence rate ($p=1.82$) in the case where only the numerical parameter $\Delta x$ is varied, with relative errors close to $0.15\%$ for the finest mesh, see Figure~\ref{fig:conv_disk}. Unsurprisingly, the convergence behaviour changes when the body thickness is simultaneously reduced with the grid size, see Figure~\ref{fig:conv_disk_2}. However, the fine mesh results presented are still within $2\%$ of the Richardson extrapolated value for $d\rightarrow 0$. We use the fine mesh for all the simulations presented hereafter.}

\subsection{Results}
\label{S_5:2}

Our results focus on a single disk acceleration, $a^*=0.5$, as the change in fluid response as a function of $a^*$ is documented in the study by Fernando et al. \cite{Fernando2020}. In addition to testing the three immersed boundary methods, we also vary the disk thickness $d$ for the BDIM and BDIM-$\sigma$ methods to test the performance for very thin bodies. We use a constant $d/\Delta x =4$ for the Direct-forcing method as the pressure BC is not enforced regardless of thickness.

\begin{figure}
    \centering
    \includegraphics[scale=1]{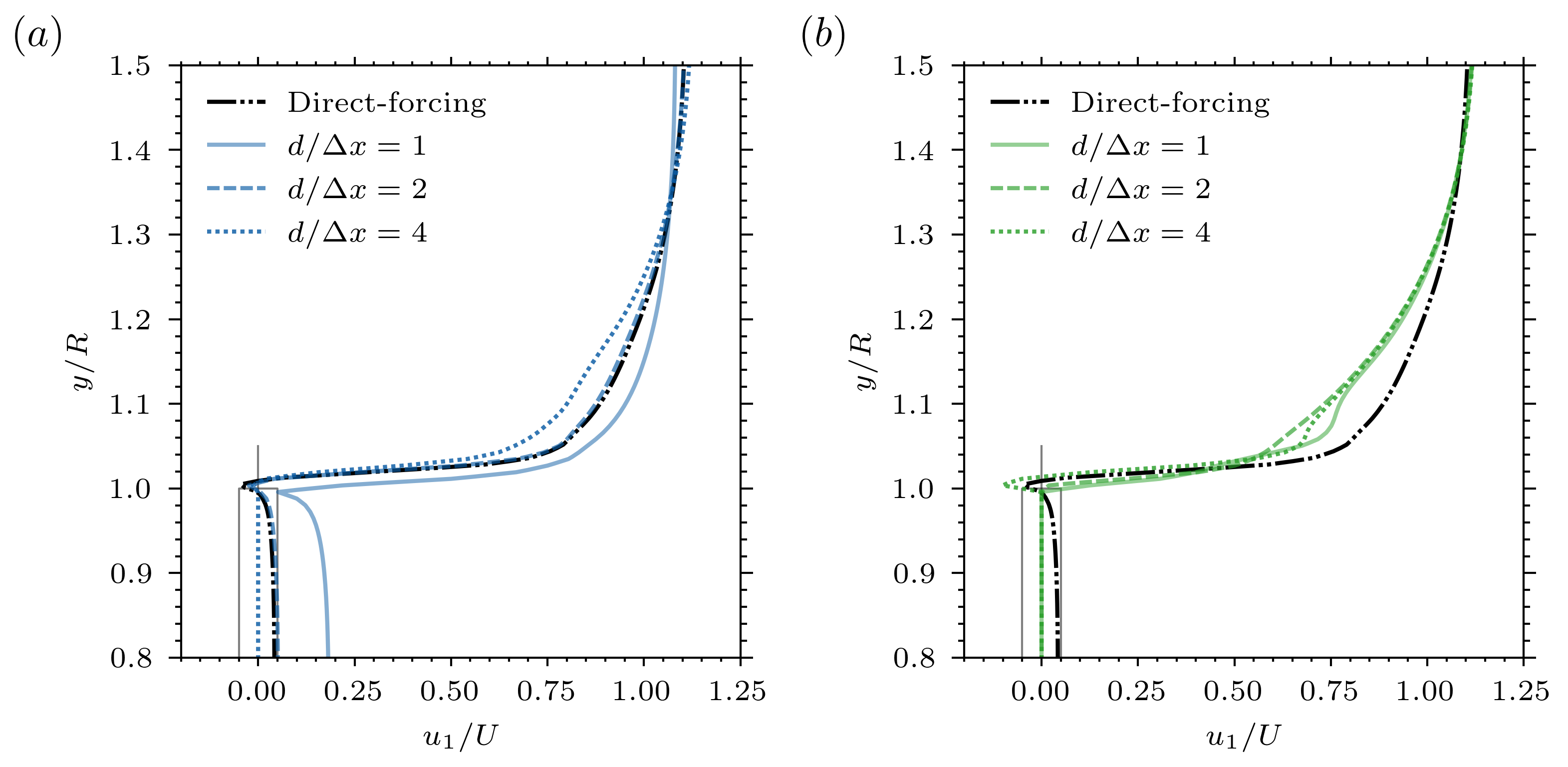}
    \caption{Axial ($u_1$) velocity profile on the top of the accelerated thin disk at a non-dimensional time $t^*=4$ for $(a)$ BDIM and Direct-forcing and $(b)$ BDIM-$\sigma$ and Direct-forcing. The thickness of the disk is exaggerated for visualisation purpose.}
    \label{fig:uprofile}
\end{figure}

Axial velocity profiles through the center-plane of the disk near the edge are shown on Figure~\ref{fig:uprofile}. The profiles are taken at $t^*=4$, which is during the steady motion phase. The solutions for Direct-forcing (with $d/\Delta x=4$) and BDIM (when $d/\Delta x<4$) are both seen to leak through the disk, violating the normal velocity condition. In contrast, BDIM-$\sigma$ always enforces the velocity BC, letting us study the influence of changing disk thickness on the flow without this modelling error. We observe differences in the velocity profiles resulting in a different shear-layer behaviour, depending on the disk thickness. Thicker disks result in separation from the front corner in addition to the back, leading to a more complex velocity profile.

\begin{figure}
    \centering
    \includegraphics[scale=1]{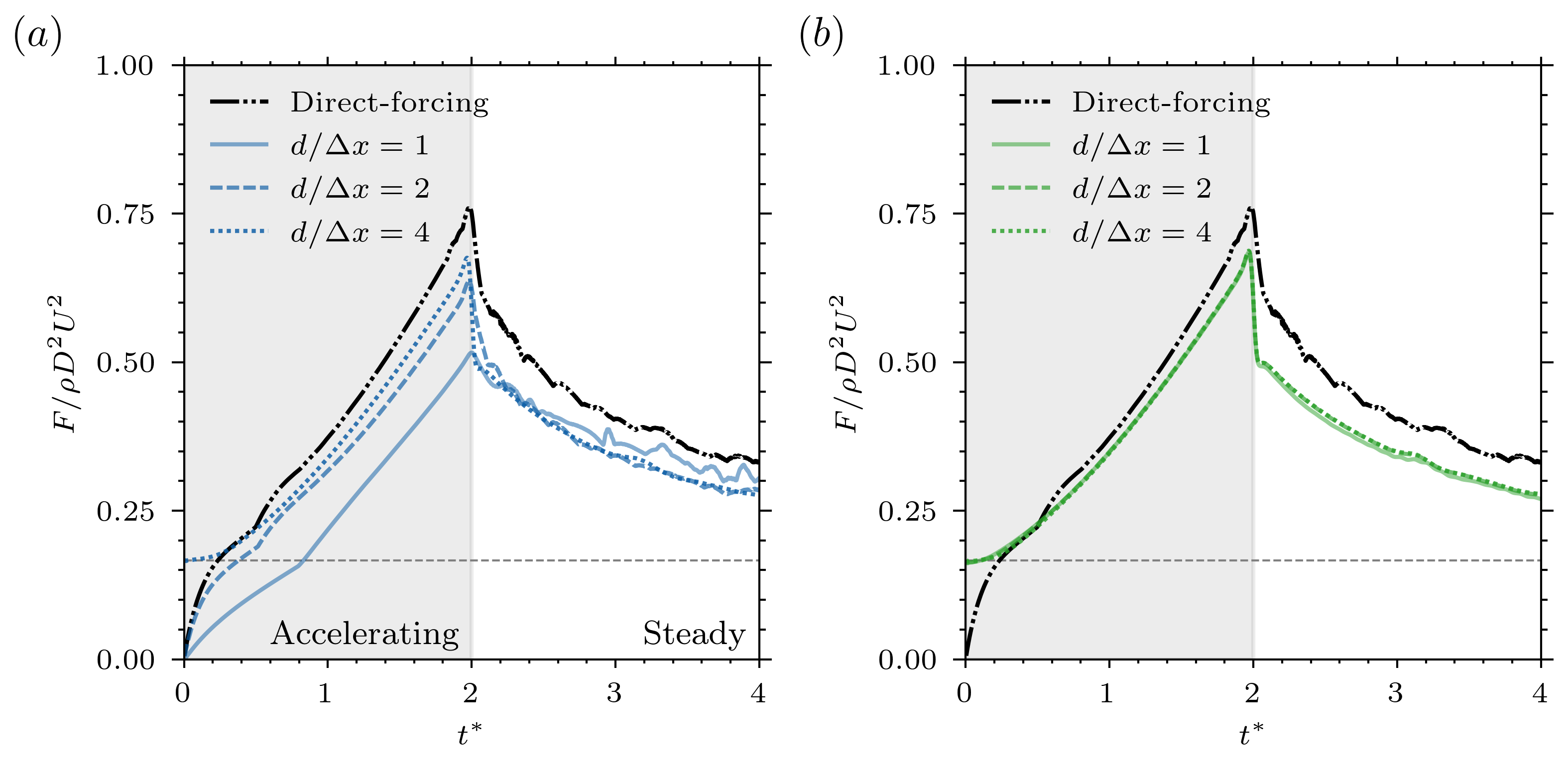}
    \caption{Normalized pressure force acting on an impulsively started disk $(a)$ BDIM and Direct-forcing, $(b)$ BDIM-$\sigma$ and Direct-forcing for $a^*=0.5$. Results are presented with various disk thicknesses (in terms of FV mesh spacing). The horizontal dashed line ($F/\rho D^2U^2=a^*/3$) represents the theoretical added-mass coefficient (only valid for the initial impulse). The accelerating ($v_b=at$) and steady phase ($v_b=U$) are represented by the gray and white shaded areas, respectively.}
\label{fig:added_mass}
\end{figure}

The resulting time history of the normalized pressure force acting on the disk can be seen on Figure~\ref{fig:added_mass}. The initial added-mass force is compared to the analytical expression for the added-mass coefficient
\begin{equation}
    C_a = \frac{F}{\rho D^3 a} = \left(\frac{F}{\rho D^2 U^2}\right)\frac{1}{a^*}=\frac 13,
\end{equation}
from potential flow theory for a thin circular disk, see \citet{brennen1982review}. This value is shown as a horizontal dashed line in Figure~\ref{fig:added_mass} and the initial force predictions for each method are summarized in Table~\ref{tab:disk_peak}. The Direct-forcing method completely misses this initial added-mass force, while BDIM is able to capture the force accurately, provided the body is sufficiently thick ($d/\Delta x\ge 4$). The new BDIM-$\sigma$ method ensures the correct pressure boundary condition is always applied and so the initial force is always captured.

\begin{table}
\centering
    \caption{Normalized initial and peak force at $U/U_{max} = 0$ and $U/U_{max} = 1$ for the different numerical methods and the experimental data. Numerical results are shown for the finest mesh with various thickness to diameter ratios $d/D$. The Direct-forcing results use $d/D=4/512$. The relative error compared to the reference solution is also shown.}
    \begin{tabular}{ll|cccc}
    Method & $d/D$ & $F_{init}$ & \footnotesize{\%Error} & $F_{peak}$ & \footnotesize{\%Error} \\
    \hline
    \multicolumn{2}{l|}{Direct-forcing \cite{Fadlun2000}} & 0.012 & \footnotesize{96.4} & 1.52 & \footnotesize{14.4}\\
    \hline
    \multirow{3}{*}{BDIM} & 1/512 & 0.001 & \footnotesize{99.6} & 1.032 & \footnotesize{22.4}\\
                          & 2/512 & 0.010 & \footnotesize{97.1} & 1.272 & \footnotesize{4.36}\\
                          & 4/512 & 0.333 & \footnotesize{0.06} & 1.351 & \footnotesize{1.61}\\
    \hline
    \multirow{3}{*}{BDIM-$\sigma$}  & 1/512 & 0.332 & \footnotesize{0.36} & 1.369 & \footnotesize{2.96}\\
                                    & 2/512 & 0.332 & \footnotesize{0.26} & 1.375 & \footnotesize{3.42}\\
                                    & 4/512 & 0.333 & \footnotesize{0.08} & 1.373 & \footnotesize{3.20}\\
    \hline
    \multirow{2}{*}{Reference} & Analytical   & 1/3.  & - & -  & -\\
    & Experimental \cite{Fernando2020,kaiser_kriegseis_rival_2020} &   -  & - &  1.33 $\pm$ 0.02 & -\\
    \end{tabular}
    \label{tab:disk_peak}
\end{table}

\begin{figure}
    \centering
    \includegraphics[scale=1]{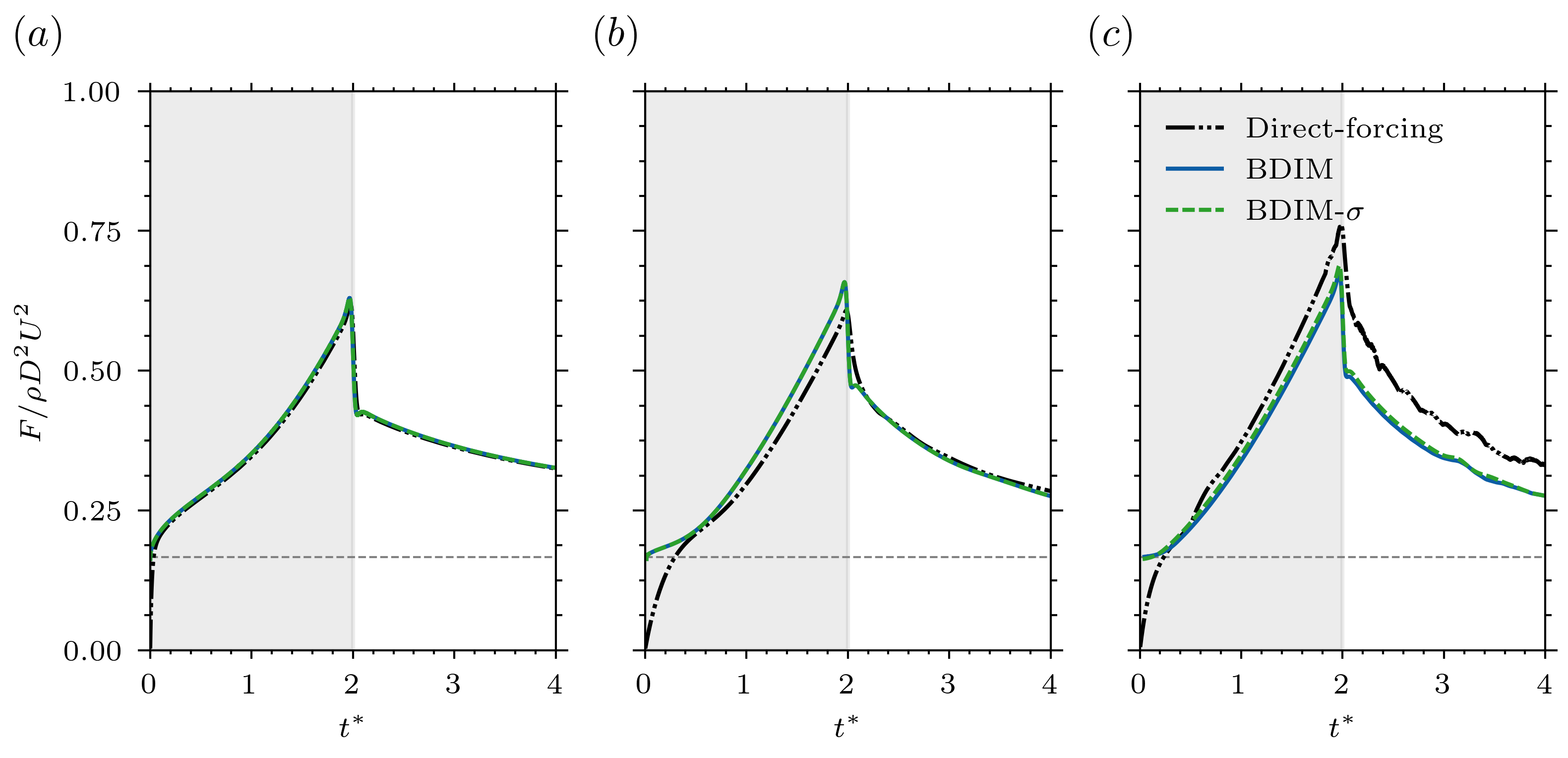}
    \caption{Normalized pressure force acting on an impulsively started disk for the BDIM, BDIM-$\sigma$, and Direct-forcing method using $d/D=4/512$ at Reynolds number, $(a)$ 100, $(b)$ 1250 and $(c)$ $125\times10^3$ for $a^*=0.5$. The horizontal dashed line ($F/\rho D^2U^2=a^*/3$) represents the theoretical added-mass coefficient (only valid for the initial impulse).}
    \label{fig:reynolds_dep}
\end{figure}

In addition to validating the initial impulse force with potential flow theory, we validate the force peak ($F_{peak}$) using published experimental data ~\citep{Fernando2020,kaiser_kriegseis_rival_2020}. The results are presented in Table~\ref{tab:disk_peak}. The experimental values are obtained for the same acceleration, but with $Re=50\times10^3$. This difference in $Re$ is not significant as \citet{Fernando2020} shows the peak forces are very weakly sensitive to the Reynolds number and we confirm this in Figure~\ref{fig:reynolds_dep}. The experimental peak force is obtained from an average of the peak forces from 125 measurements with a disk of $d/D=0.01$. The standard deviation of $1.3\%$ is used as an estimate of the experimental uncertainty in this value. As before, Direct-forcing method has the largest error, although now the lack of pressure boundary condition has resulted in an \textit{over}-estimate of the pressure force which continues through the steady velocity phase ($t^*>2$). \textcolor{black}{Table~\ref{tab:disk_peak} shows the classic BDIM method approaches (and overshoots) the peak experimental drag as the thickness is increased. However, classic BDIM cannot enforce the pressure boundary condition exactly when $d/D=1/512$ (corresponding to $d=2\Delta x$) which results in more than 20\% error. 
In contrast, BDIM-$\sigma$ always enforces the pressure boundary condition and therefore gives an essentially constant force prediction within around 3\% of the experiments even as $d\rightarrow 0$.}

Finally, we verify the arguments in Section~\ref{bdimsigma} and show the increased importance of the pressure boundary condition with increased $Re$ (Figure~\ref{fig:reynolds_dep}). The BDIM and BDIM-$\sigma$ solution using $d/D=4/512$ are indistinguishable and show very little change in the force profile as the Reynolds number is increased from $Re=100$ to $1250$ to $125000$, especially during the acceleration phase ($t^*<2)$, in agreement with the experimental findings \cite{Fernando2020,kaiser_kriegseis_rival_2020}. In contrast, the Direct-forcing results only match BDIM and BDIM-$\sigma$ for $Re=100$, with large force errors at higher $Re$. The leakage observed and incorrect force predictions are purely due to the pressure field being computed without proper boundary conditions on the body. This results in a non-zero pressure flux across the disk during the projection step, and this error is exaggerated as the flow gradients near the boundary increase, i.e. at higher $Re$. The only way to avoid this issue for all $Re$ is to adjust the Poisson equation for the pressure such that the Neumann boundary condition is imposed explicitly.

\section{Application: Flapping Insect Wing}
\label{S:Wing}

As a final example, we consider the flow induced by a flapping insect wing, demonstrating the performance of immersed boundary methods when modelling moving three-dimensional thin structures. The geometry is a simplified representation of a \emph{drosophila} wing (a common fly), commonly modeled as an ellipsoid~\citep{Wang2004, Sane2001, Bos2013, Zheng2020, Bos2008} defined by the following equation:
\begin{equation}
    \frac{x^2}{a^2} + \frac{y^2}{b^2} + \frac{z^2}{c^2} = 1,
\end{equation}
where $a=0.5$, $b=0.05$ and $c=1$. However, to highlight the ability of BDIM-$\sigma$ to approach the true wing's geometry, we model the wing as an elliptically-shaped and initially flat shell ($i.e.$ a plate) with constant minimal thickness $d$. Of course, the two representations match in the limit of $b\rightarrow 0$ and $d\rightarrow 0$, as the wing becomes an ellipsoidal flat plate.

\begin{figure}
    \centering
    \includegraphics[width=\textwidth]{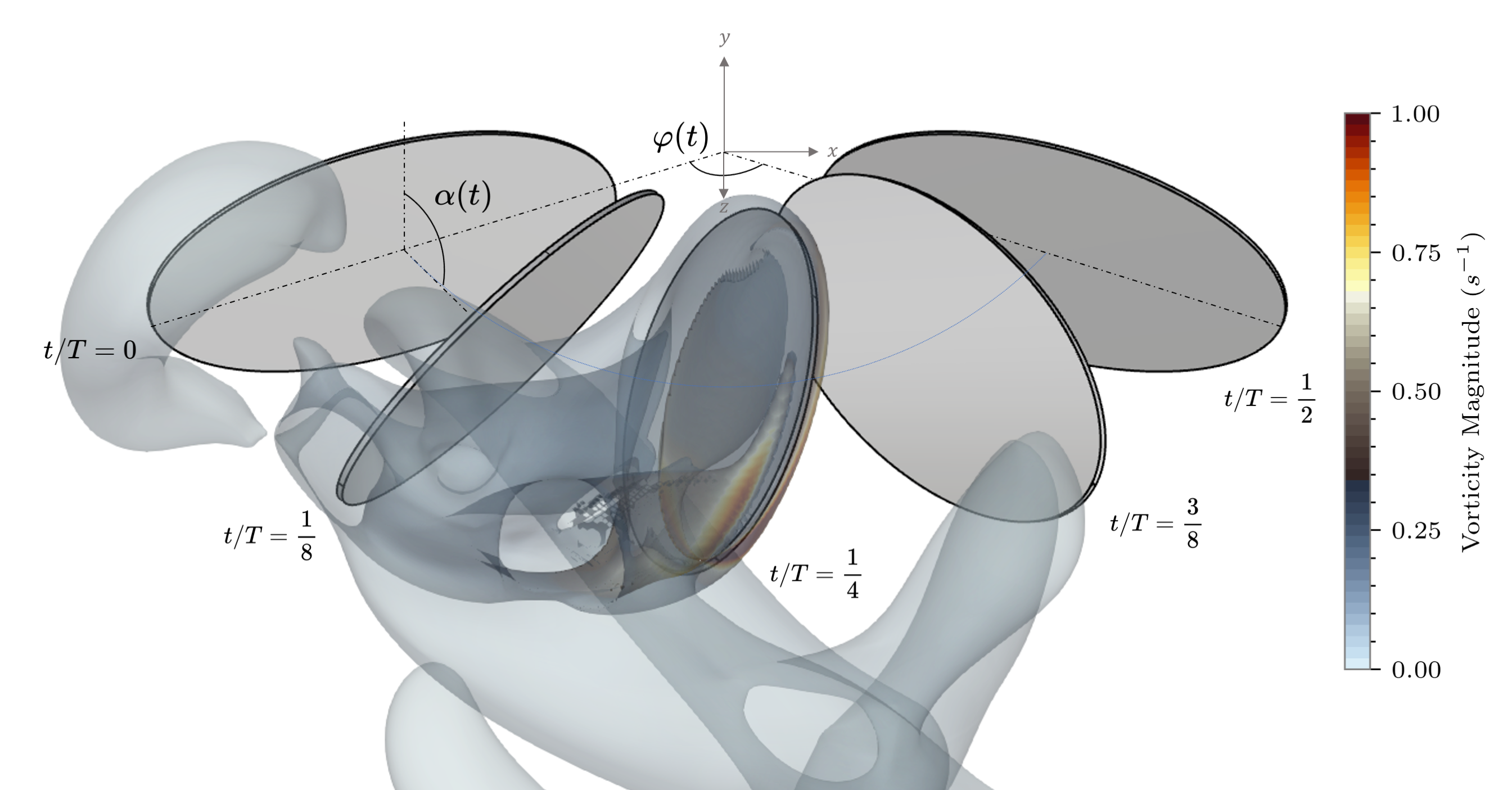}
    \caption{Elliptical plate wing at different time during an oscillatory period. Motion kinematics are: $f_r=0.5$, $A_\theta=0$, $A_\varphi=0.35\pi$ and $A_\alpha=\pi/4$. The blue line represents the motion of the center of the wing. The vortical structures are shown as iso-surface of $\lambda_2$-criterion ($\lambda_2 U^2/D^2=-8$) colored with vorticity magnitude ($\omega U/D$), shown only for $t/T=1/4$. Animations of the above figure are provided as supplementary material.}
    \label{fig:ellipse}
\end{figure}

The wing's motion is completely described by the three following Euler angles
\begin{subequations}
\begin{align}
    \varphi(t) &= A_\varphi\cos(2\pi f_r t),\\
    \theta(t) &= A_\theta\sin(2\pi f_\theta t),\\
    \alpha(t) &= \frac{\pi}{2} - A_\alpha\sin(2\pi f_r t + \xi).
\end{align}
\end{subequations}
We set the flapping amplitude to $A_\varphi=0.35\pi$ and the angle of attack amplitude to $A_\alpha=\pi/4$, see Figure~\ref{fig:ellipse}, to replicate the numerical results presented in~\cite{Zheng2020}. The deviation amplitude, $A_\theta$ is set to zero such that the flapping motion is constrained to the $x-z$ plane. The flapping and angle of attack frequencies are taken as $f_r=0.5$ $Hz$. The deviation of the angle of attack to the flapping angle is set to zero $\xi=0$. These kinematics result in a Rosby number of $Ro\equiv R_{tip}/\bar{c}=3.2$ and a Reynolds number of $Re\equiv\bar{c}U/\nu=100$, where $R_{tip}$ is the distance from the instantaneous center of motion (in this case, the origin) to the tip of the wing and $\bar c$ is the mean chord of the wing. The reference velocity is defined as $U=f_rR_g\int_{0}^{1/f_r}\sqrt{\dot{\varphi}^2+\dot{\theta}^2}\diff t$, and the radius of gyration $R_g = \sqrt{\frac{1}{S}\int_{0}^{R}cr^2 \diff r}$.

\subsection{Numerical Method}

\begin{figure}
    \centering\includegraphics[width=\textwidth]{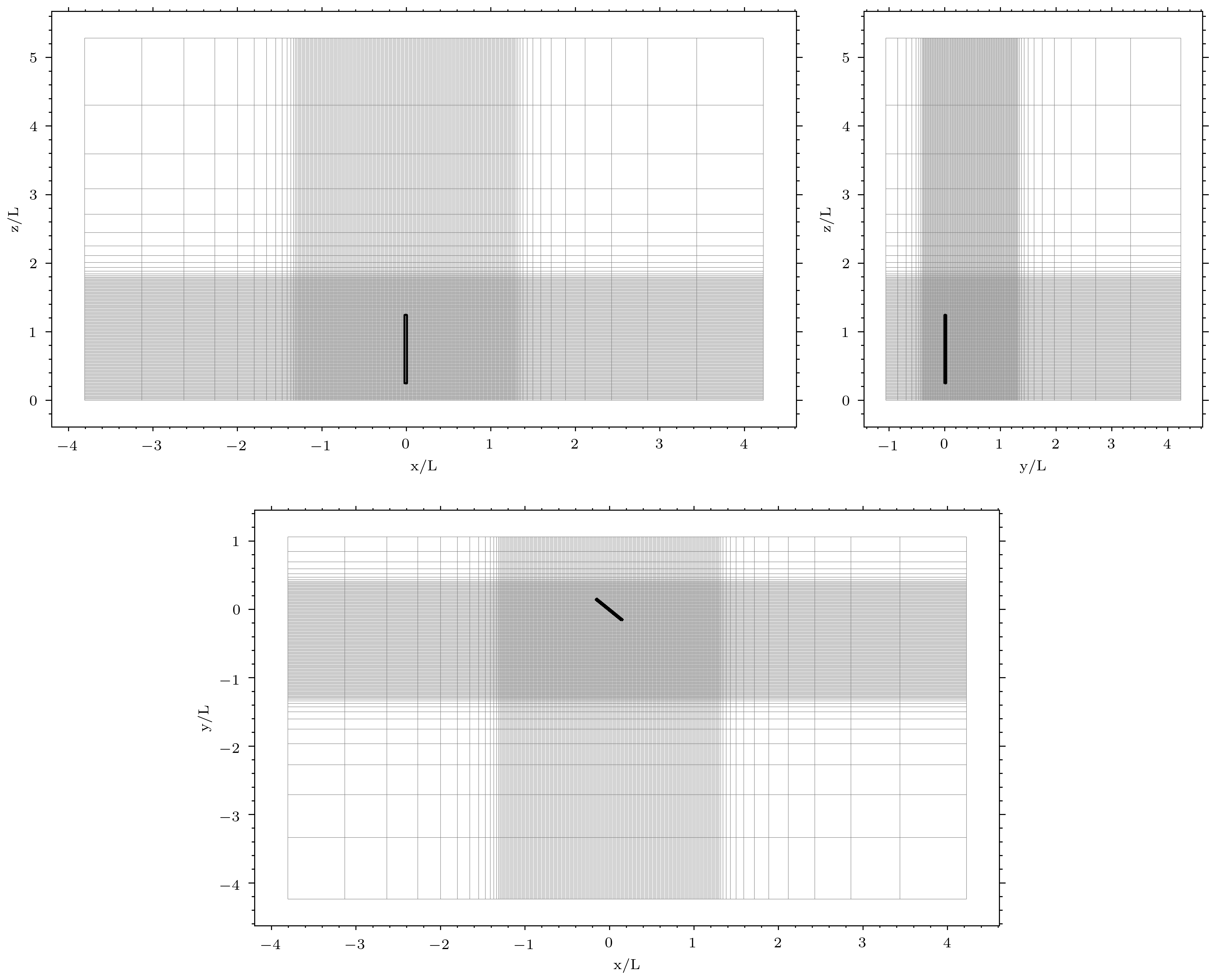}
    \caption{First-angle projection of the finest grid used for the ellipsoidal wing case. The wing is shown at $t/T=0.25$. For clarity every two cells are shown in each direction.}
    \label{fig:grid_wing}
\end{figure}

\textcolor{black}{A uniform grid region of $[2.6\times1.7\times1.8]$ wing span is used to properly capture the unsteady flow structures around the wing, see Figure~\ref{fig:grid_wing}. Grid stretching is used to fill the rest of the domain until it reaches the total size of $[8\times5.5\times5]$ wing span. A zero normal flux condition is applied on all exterior boundary of the domain. The no-slip boundary condition is applied to the immersed body. }

Convergence is assessed using the mean lift coefficient over the last four periods (good periodicity is found for the last four period of motion of our system, see Figure~\ref{fig:conv_wing}d). We vary the resolution while using the thinnest possible body. The convergence rate with thickness is found to be $O((d/D)^{1.55})$ with relative error to the Richardson extrapolated value of $2.7\%$ for the finest mesh, and so this is the mesh used in the results. As with the accelerating disk case, the convergence rate is less than 2 because the thickness is simultaneously reducing with $\Delta x$.

\begin{figure}
    \centering\includegraphics[width=\textwidth]{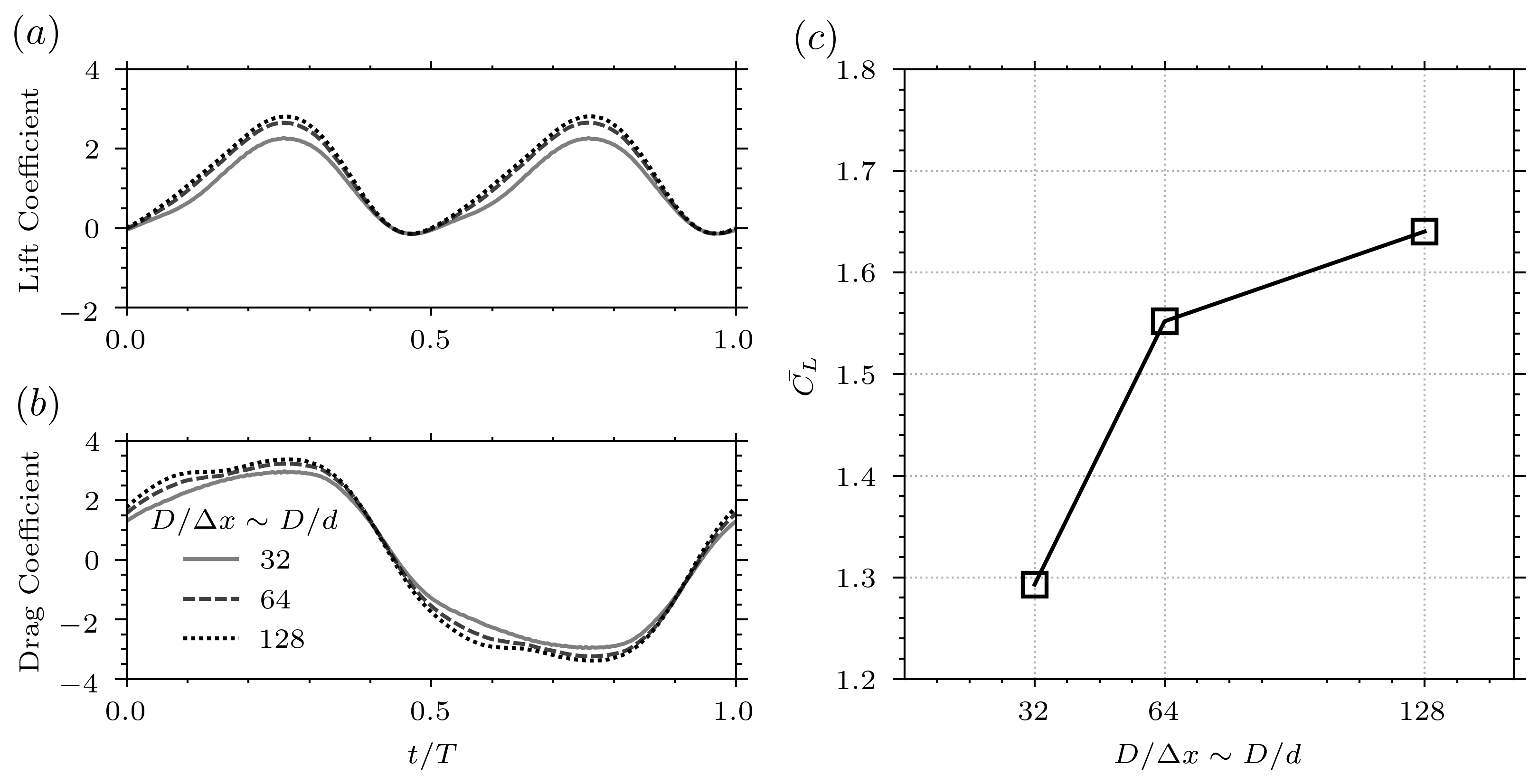}
    \caption{Geometrical sensitivity study of the $(a)$ lift and $(b)$ drag coefficient for the wing with the minimum body thickness $d=(1+\sqrt{3})\Delta x$. $(c)$ convergence of the mean lift coefficient for varying resolution and thicknesses.}
    \label{fig:convergence_wing}
\end{figure}

\subsection{Results}

Results are presented in terms of time dependent lift and drag coefficients
\begin{equation}
    C_L = \frac{F_y}{\frac{1}{2}\rho U^2S}\;,  \qquad C_D = \frac{F_x\cos(\varphi)-F_z\sin(\varphi)}{\frac{1}{2}\rho U^2S}\;,
\end{equation}
where $S$ is the surface area of the ellipsoidal wing and $F_i$ is the $i^{th}$ component of the instantaneous force acting on the wing. \textcolor{black}{We model the wing with the same thickness $d=(1+\sqrt{3})$ for all the Immersed Boundary methods.}

\begin{figure}
    \centering
    \includegraphics[scale=1.]{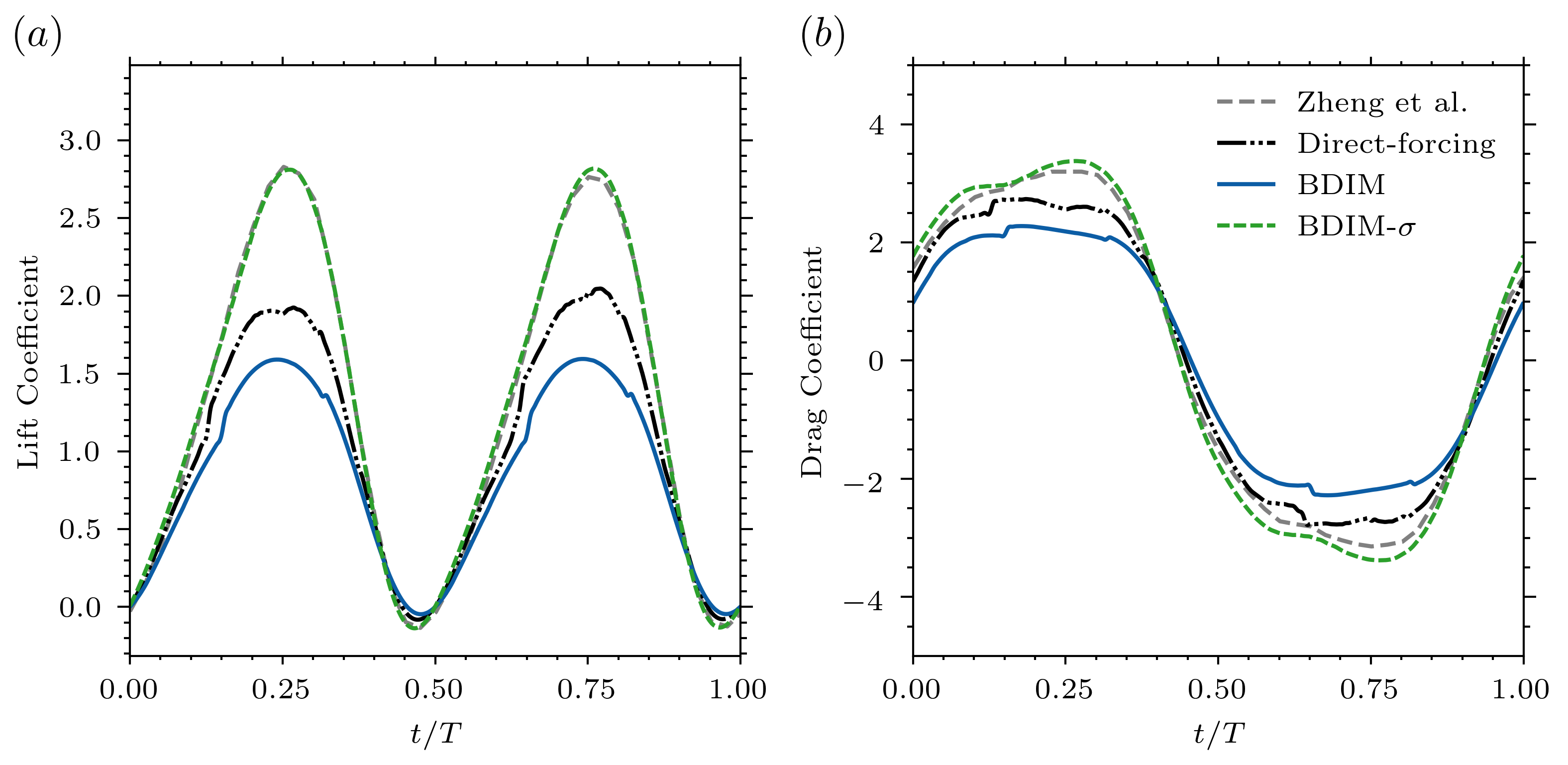}
    \caption{Time history of the $(a)$ lift and $(b)$ drag coefficient for the wing. Results are presented for a single period after the motion become independent of the initial conditions.}
    \label{fig:ellipse_forces_1}
\end{figure}

\begin{figure}
    \centering
    \includegraphics[scale=1.]{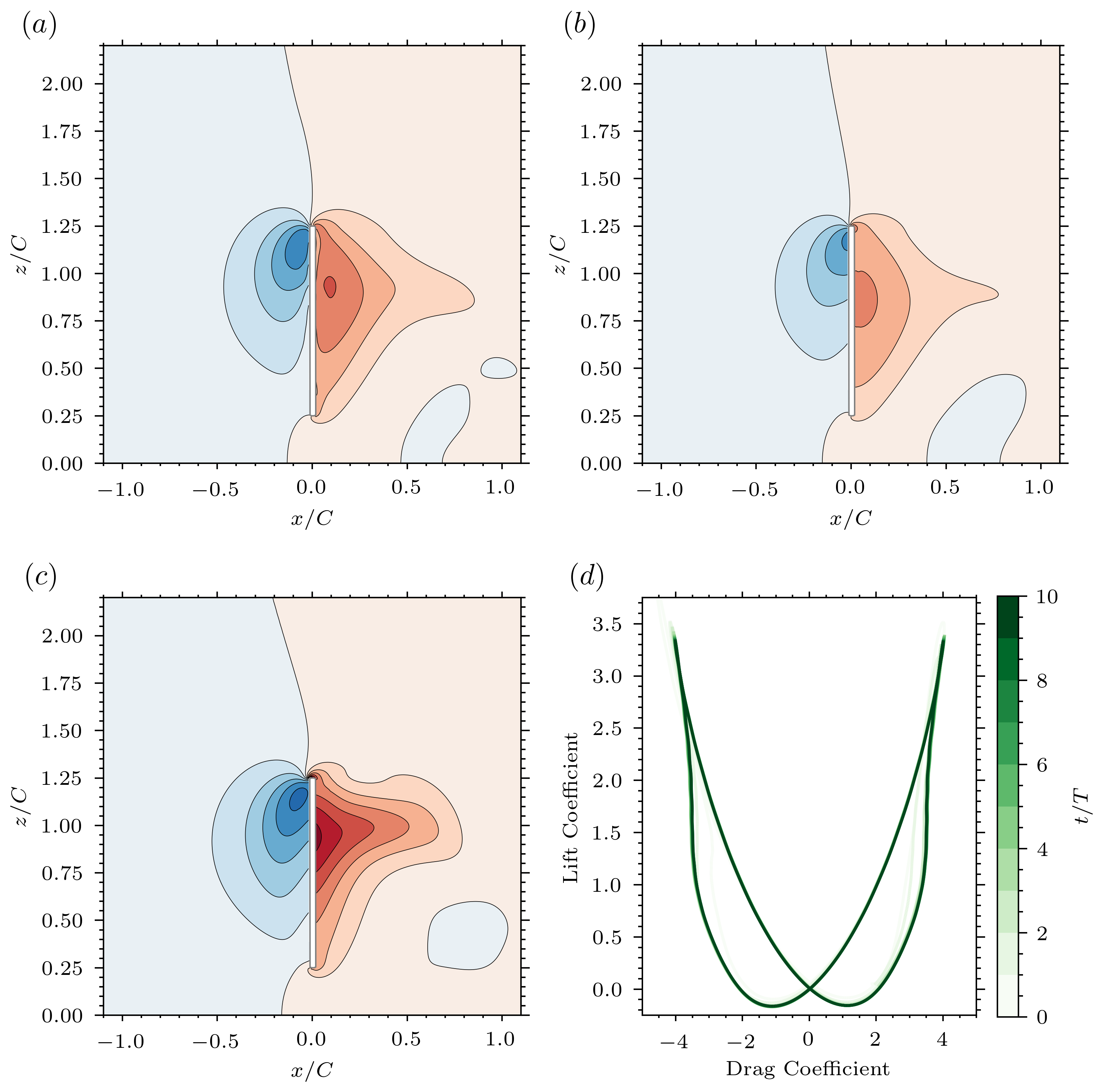}
    \caption{Pressure contour at $t/T=0.25$ for the ellipsoidal wing with a thickness $(1+\sqrt{3})\Delta x$ for $(a)$ the Direct-forcing method $(b)$ BDIM and $(c)$ BDIM-$\sigma$. The 15 contour levels are evenly spaced between $c_p = \pm 7.0$. $(d)$ Phase portrait of the wing forces for the BDIM-$\sigma$ method. Clear periodicity is found for $t/T>6$.}
    \label{fig:conv_wing}
\end{figure}

We include additional numerical results of the flow around an ellipsoidal wing obtained in \citet{Zheng2020}. \textcolor{black}{Despite the fact that \cite{Zheng2020} models the wing as a thin ellipsoid instead of an elliptical plate the lift and drag coefficient results are very comparable to BDIM-$\sigma$. The Direct-forcing and BDIM results shows a large under-prediction of the forces. Figure~\ref{fig:conv_wing}a-c shows that this large under-prediction is due to a different pressure field around the body. In the Direct-forcing method this pressure field is the result of the non-existant pressure boundary condition applied to the Poisson equation. Because the body is very thin, the standard BDIM method does not properly impose the pressure boundary condition either, which results in a leakage of fluid across the interface. This issue is resolved with our new method which allows the simulation of very thin membrane wings, bringing our simulation closer to real-life insects wings than the simplified ellipsoidal wing used in previous studies.} As detailed in \ref{A:4}, the reduced smoothing width $\epsilon$ used in BDIM-$\sigma$ induces a small high-frequency numerical noise in the pressure force's temporal-spectra. However, the magnitude of the noise is still less than the Direct-forcing method, six orders of magnitude smaller than the low frequency physical signal, and does not corrupt the spatial pressure field, see Figure~\ref{fig:conv_wing}c.

\begin{figure}
    \centering
    \includegraphics[scale=1.]{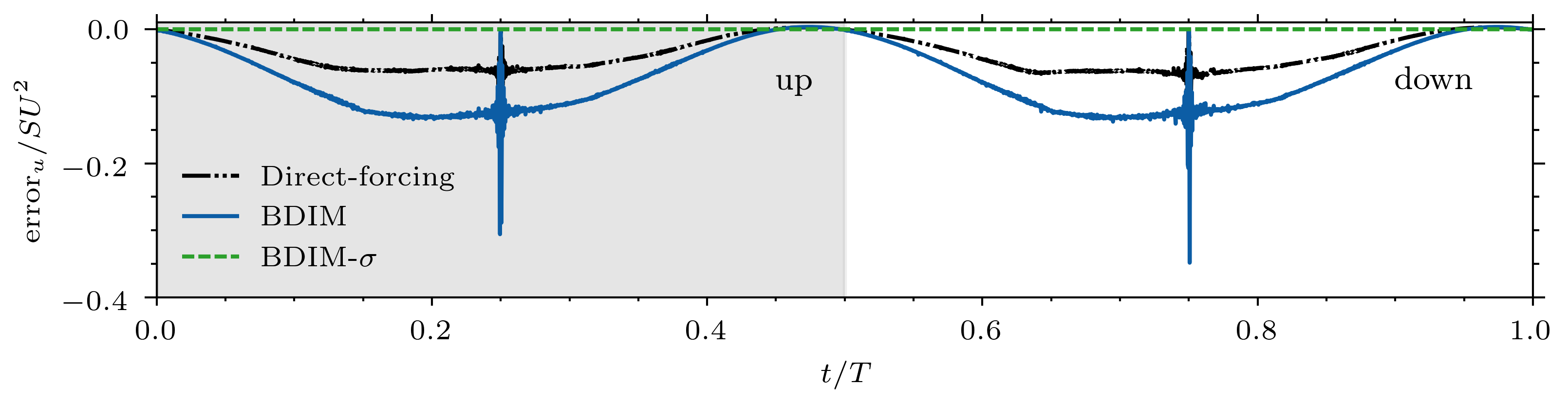}
    \caption{Time variation of the velocity error in Direct-forcing method, BDIM and the BDIM-$\sigma$ for a single motion period. The gray in white shaded areas represent the up and down stroke, respectively. The surface integral over the mid-plane of the wing is normalized by the product of wing area and velocity magnitude squared $S\,U^2$. The $L_\infty$-norm of the velocity error for the BDIM-$\sigma$ for the period shown is $2.261\times10^{-7}$.}
    \label{fig:velocity_error}
\end{figure}

Because the normal pressure gradient close to the immersed body doesn't match the local body acceleration, the Direct-forcing method and BDIM results in an error in the fluid velocity inside the body. We measure the error in the momentum flux of fluid through the wing's mid-plane, defined as
\begin{equation}
    \text{error}_u \equiv \oint_{\sigma} (\vec{u}-\vec{v}_b)\cdot\vec{v}_b \diff{x_B},
\end{equation}
where $\vec{v}_b$ is the body velocity, and $\sigma$ is the surface representing the mid-plane of the ellipsoidal wing. \textcolor{black}{We approximate this surface integral numerically by integrating over all cells that are located on the mid-plane of the body. In Direct-forcing methods, this error is only introduced in the projection step, and is not present in the intermediate velocity field, whereas it is already present in the intermediate velocity field in BDIM}. The time variation of this error is shown on Figure~\ref{fig:velocity_error}. The large negative error observed represents a deficit in the local fluid velocity inside the body. This means that the forcing generated by the immersed surface is not properly radiated onto the fluid field, in agreement with the under-predicted lift and drag forces discussed above. A reversal of the sign of the error is observed just before the wing reaches the maximum flapping amplitude ($t/T=0.5$ or $t/T=1$).

\section{Conclusions}

\textcolor{black}{This manuscript provides analytical and numerical demonstration that most immersed boundary methods violate the pressure boundary condition for thin dynamic surfaces, resulting in a violation of the velocity boundary condition and erroneous flow fields and pressures forces. With the help of an illustrative one-dimensional problem, we show that the treatment of the pressure boundary condition in immersed boundary method is closely linked to the accuracy of the final velocity field, and that the magnitude of this error increases with Reynolds number. An analysis of the discretized pressure Poisson equation shows that in order to enforce the pressure boundary condition correctly the immersed boundary method must either solve an augmented system or modify the Poisson matrix coefficients such that the solution may be discontinuous across the boundary.}

This analysis allows us to specify an extension of the original BDIM method to thin dynamic surface called BDIM-$\sigma$. This new method adjusts the signed distance function to maintain a minimal body thickness that ensures proper imposition of the Neumann condition onto the body. The method relies on the \text{slender-body} notation to describe the mid-plane of the body with prescribed thickness. We note that this notation is commonly used to describe the kinematic of thin shells (see \cite{Chen2014,Duong2017}, for examples) and therefore provides a link for possible use of this modified method for fluid-structure interaction problems involving thin flexible structures.

\textcolor{black}{Challenging three-dimensional test cases demonstrate that the new BDIM-$\sigma$ approach outperforms the Direct-forcing method and the original BDIM approach when dealing with thin dynamic bodies dominated by pressure forces, as is typical in many intermediate and high Reynolds number fluid-structure interaction problems. In particular, BDIM-$\sigma$ always enforces the velocity and pressure boundary conditions on the surfaces regardless of resolution, ensuring accurate flow and force predictions compared to analytic solutions, previous high-resolution numerical studies, and experimental results.}

The ability of the new BDIM-$\sigma$ approach to deal with thin dynamic surfaces opens a wide range of exciting fluid-structure interaction applications such as sails, flexible insect wings, swimming fins, inflatable structures and many others. We believe that this approach will allow accurate and efficient flow simulations around highly dynamic non-linear shells and membranes.

\section{Acknowledgements}

The authors would like to acknowledge financial support from the EPSRC Centre for Doctoral Training in Next Generation Computational Modelling grant EP/L015382/1, and the use of the IRIDIS High Performance Computing Facility and associated support services at the University of Southampton. The authors gratefully acknowledge Bernat Font and Artur K. Lidtke for their helpful comments on an early version of this manuscript. \textcolor{black}{Finally, we would like to acknowledge the reviewers that made excellent comments and suggestions that, we believe, helped greatly improve this manuscript.}

\appendix

\section{Taylor-Green Vortex}
\label{A:1}
\textcolor{black}{We demonstrate the accuracy of the flow solver \emph{without} an immersed body by simulating the decay of a periodic vortex array on a domain $x,y\in[0, 1]$. The decay of the velocity and pressure field is governed by
\begin{subequations}
\begin{align}
    u(x,y,t) &= \sin(k_x x)\cos(k_y y)\text{e}^{-(k_x^2+k_y^2)\nu t}\\
    v(x,y,t) &= -\cos(k_x x)\sin(k_y y)\text{e}^{-(k_x^2+k_y^2)\nu t}\\
    p(x,y,t) &= \frac{1}{4}\left(\cos(2k_x x)+\cos(2k_y y)\right)\text{e}^{-(k_x^2+k_y^2)\nu t}
\end{align}
\end{subequations}
with $k_x=k_y=2\pi$ and $\nu=0.001$. This analytical solution to the Navier-Stokes equation serves as initial condition for the simulations and exact solution from which we measure the error using the $L$-2 norm (see Equation~\ref{eq:L2}). We use four distinct mesh of size $N^2$ with $N\in[64,128,256,512]$ and compute the error at $t=1s$ and compared to the analytical solution, results are presented on Figure~\ref{fig:tgv_convergence}a. Higher-than second-order convergence is found for both the pressure and velocity field, demonstrating the spatial accuracy of our solver.  
\begin{figure}
    \centering\includegraphics[scale=1.]{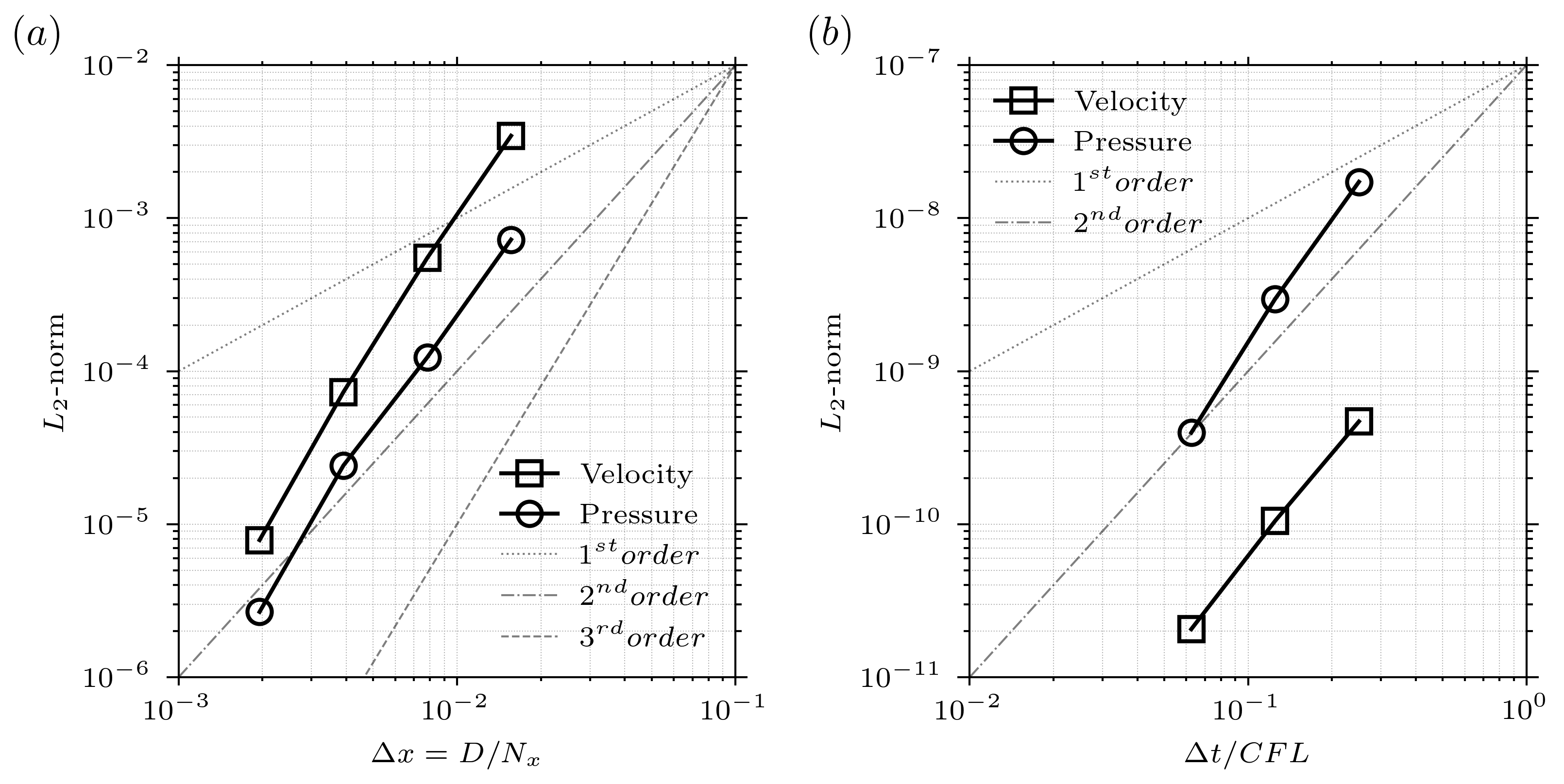}
    \caption{$(a)$ Spatial and $(b)$ Temporal convergence of the flow solver on the 2D Taylor-Green vortex.}
    \label{fig:tgv_convergence}
\end{figure}
The temporal convergence is assessed using the finest mesh ($512^2$) to keep spatial discretization error to a minimum and by gradually halving the time-step/CFL ratio with $\Delta t/CFL\in[2,4,8,16]$. Although the spatial truncation error is small at this resolution ($10^{-5}$), it is orders of magnitude larger than the temporal truncation error. This does not allow to compare the solution obtained with various time-steps to the analytical solution. As such, we use the numerical simulation obtained with $\Delta t/CFL=1/16$ as the \emph{exact} solution and compute the error with reference to this result. Second-order convergence is found for the velocity field, with higher-than second-order convergence in the pressure field, see Figure~\ref{fig:tgv_convergence}b.}

\section{Force Spectra}
\label{A:4}

\begin{figure}
    \centering\includegraphics[scale=1.]{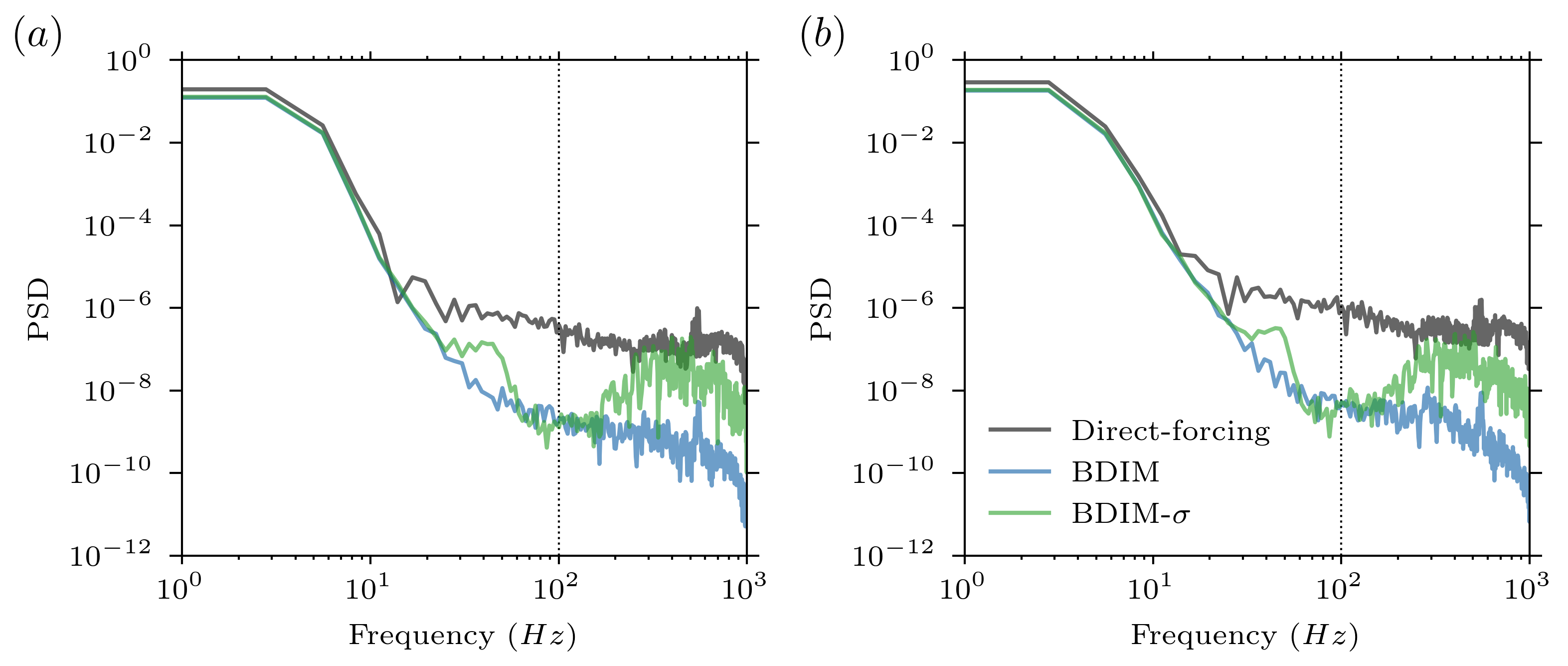}
    \caption{Spectral density of the time-dependent lift $(a)$ and drag $(b)$ coefficient. We use a simple high-frequency filter to remove the noise above 100 $Hz$.}
    \label{fig:filter}
\end{figure}

The spectral content of the lift and drag coefficient for the different methods are shown in Figure~\ref{fig:filter}. To remove the temporal noise in the force integral over the surface of the body, we use a high-frequency filter with a cuttoff frequency of $100$ $Hz$.

\begin{equation}
    \tilde C_L(f) = \begin{cases}
    \tilde C_L(f) &\text{ if }  f < f_{\text{cutoff}},\\
    0. &\text{ else. }
    \end{cases}
\end{equation}

\bibliographystyle{elsarticle-num-names}
\bibliography{sample.bib}







\end{document}